\documentclass[a4paper,11pt]{article}
\pdfoutput=1 

\usepackage{jheppub} 

\usepackage[T1]{fontenc} 

\usepackage{amssymb}

\usepackage[utf8]{inputenc} 
\usepackage{graphicx}

\usepackage{braket}
\usepackage[vcentermath]{youngtab}

 \usepackage{fancyhdr}
\usepackage{bbm}
\usepackage{mathrsfs}
\usepackage[T1]{fontenc}
\usepackage{setspace}
\usepackage{amsfonts}
\usepackage{amssymb}
\usepackage{amsmath}
\usepackage{epsfig}
\usepackage{latexsym}
\usepackage{color}
\usepackage{nicefrac}
 \usepackage{slashed}
 \usepackage{multirow}
 \usepackage{comment}
 \usepackage{soul}
 \usepackage{hyperref}
\usepackage{slashed}
\usepackage{simpler-wick}
\usepackage{array}
\usepackage{tabu}
\usepackage{tcolorbox}
\usepackage{tabularx}

\newcommand{\ea}[1]{
\begin{align}
#1
\end{align}
}

\newcommand{\tad}[1]{\textcolor{blue}{#1} }
\newcommand{\beq}{\begin{eqnarray}}
\newcommand{\eeq}{\end{eqnarray}}

\newcommand{\p}{{\cal P}\exp}

\newcommand{\bmp}{\noindent\begin{minipage}{16cm}}
\newcommand{\emp}{\end{minipage}\vskip 7mm} 

\newcommand   \cO {\mathcal{O}}

\renewcommand \l  {\lambda}

\newcommand{\Tr}{\text{Tr}}
\newcommand*{\del}{\mathop{\mathrm{{}\partial}}\mathopen{}}

\newcommand{\ee}{\end{equation}}
\newcommand{\be}{\begin{equation}}
\newcommand{\bea}{\begin{align}}
\newcommand{\eea}{\end{alig}}

\renewcommand{\d}{\textrm{d}}

\def\lsim{\mathrel{\rlap{\lower4pt\hbox{\hskip1pt$\sim$}}
    \raise1pt\hbox{$<$}}}                
\def\gsim{\mathrel{\rlap{\lower4pt\hbox{\hskip1pt$\sim$}}
    \raise1pt\hbox{$>$}}}                
    
    \def\be{\begin{equation}}
\def\ee{\end{equation}}
\def\ba{\begin{eqnarray}}
\def\ea{\end{eqnarray}}

\def\del{\partial}

\def\a{\alpha}

\def\b{\beta}

\def\g{\gamma}
\def\G{\Gamma}
\def\d{\delta}
\def\D{\Delta}

\def\p{\pi}

\def\m{\mu}
\def\n{\nu}

\def\l{\lambda}

\def\s{\sigma}

\def\qq{\qquad}

\def\IR{\relax{\rm I\kern-.18em R}}

\def\IR{\relax{\rm I\kern-.18em R}}
\def\IL{\relax{\rm I\kern-.18em L}}

\def\inv{^{\raise.15ex\hbox{${\scriptscriptstyle -}$}\kern-.05em 1}}

\def\cO{{\cal O}}

\def\Tr{{\rm Tr}}

\newcommand{\red}[1]{\textcolor{red}{#1} }

\title{\boldmath \centering Infinite order results \\ for \\ Charged sectors of the Standard Model}

\author[a]{Oleg Antipin,}
\author[b]{Jahmall Bersini,}
\author[c] {Pantelis Panopoulos,}
\author[d,e,f]{Francesco Sannino,}
\author[g] {Zhi-Wei Wang}

\affiliation[a]{Rudjer Boskovic Institute, Division of Theoretical Physics, Bijeni\v cka 54, 10000 Zagreb, Croatia}

\affiliation[b]{Kavli IPMU (WPI), UTIAS, The University of Tokyo, Kashiwa, Chiba 277-8583, Japan}

\affiliation[c]{Asia Pacific Center for Theoretical Physics, Pohang, 37673, Korea}

\affiliation[d]{Quantum  Theory Center ($\hbar$QTC) at IMADA and D-IAS, Southern Denmark Univ., Campusvej 55, 5230 Odense M, Denmark}

\affiliation[e]{Dipartimento di Fisica ``E. Pancini" and INFN sezione di NapoliUniversità di Napoli Federico II, via Cintia, 80126 Napoli, Italy} 
\affiliation[f]{Scuola Superiore Meridionale, Largo S. Marcellino, 10, 80138 Napoli NA, Italy}

\affiliation[g]{School of Physics, The University of Electronic Science and Technology of China, 88 Tian-run Road, Chengdu, China}

\emailAdd{oantipin@irb.hr}
\emailAdd{jahmall.bersini@ipmu.jp}
\emailAdd{pantelis.panopoulos@apctp.org}
\emailAdd{sannino@qtc.sdu.dk}
\emailAdd{ zhiwei.wang@uestc.edu.cn}
\abstract
{
We determine anomalous dimensions of a family of fixed hypercharge  operators in the Standard Model featuring the general Cabibbo-Kobayashi-Maskawa structure.  The results are obtained at infinite orders in the couplings and to leading and subleading orders in the charge.  The computed anomalous dimensions are shown to agree with the maximum known order in perturbation theory. We further show that the large hypercharge sector of the Standard Model is characterised by a non-Abelian vector condensation phase.   
 
}

\arxivnumber{}
\begin{document}
\maketitle
\flushbottom

\section{Introduction}

The Standard Model (SM) of particle interactions is currently the most successful theory of Nature. Therefore, testing it via direct comparison with experiments constantly requires reducing theoretical uncertainty by resorting to higher order computations. Additionally, certain sectors of the SM have not yet been explored enough both theoretically and experimentally. These include for example the physics stemming from families of composite operators featuring multiple Higgses. 

To inch forward in this direction, in this work we show how to renormalize the family of lowest-lying Higgs operators with fixed hypercharge $Q$ by determining their anomalous dimensions to infinite orders in the SM couplings and leading and subleading orders in $Q$. This will be achieved by exploiting and extending a bag of tools related to the semiclassical approach known as the large charge expansion \cite{Hellerman:2015nra, Monin:2016jmo, Badel:2019oxl, Badel:2019khk, Orlando:2019hte, Orlando:2019skh, Arias-Tamargo:2019xld,  Gaume:2020bmp, Orlando:2020yii, Antipin:2020abu, Antipin:2020rdw, Antipin:2022dsm, Jack:2020wvs, Jack:2021lja, Cuomo:2022kio, Cuomo:2020rgt, Dondi:2022zna, Antipin:2022naw, Hellerman:2021qzz, Antipin:2022hfe, Hellerman:2023myh, Caetano:2023zwe, Banerjee:2017fcx, Jafferis:2017zna}. The leading order in $Q$ is given in \eqref{LEAD}  and at the next-to-leading in \eqref{NLOFERM}, \eqref{SUMFERM}, \eqref{NLOBOS}, and \eqref{SUMBOS}. For the reader's benefit and to keep the paper self-contained we review the approach in the next section. 

The analysis is semiclassical in nature \cite{Hellerman:2015nra, Monin:2016jmo, Badel:2019oxl} and determines the aforementioned anomalous dimensions via a saddle-point expansion around a nontrivial vacuum. In the case of the SM, the relevant classical trajectory is spatially homogeneous but anisotropic and defines a vector condensation phase characterized by a nonvanishing expectation value for both the Higgs field and the electroweak gauge bosons. This is in contrast with most previous applications of the semiclassical large charge approach where the ground state was an isotropic superfluid \cite{Hellerman:2015nra, Monin:2016jmo, Badel:2019oxl, Badel:2019khk, Gaume:2020bmp, Antipin:2020abu, Antipin:2020rdw, Antipin:2022naw, Jack:2020wvs, Jack:2021lja}.

The Standard Model Lagrangian, notation and conventions are presented in Sec.~\ref{StandardModel}. The identification of the family of Higgs composite operators with fixed hypercharge is discussed in Sec.~\ref{OID} while the leading and subleading corrections in the charge $Q$ are determined respectively in Sec.~\ref{LO} and  Sec.~\ref{NLO}. We show in Sec.~\ref{disco} that our anomalous dimensions reproduce perturbative results.  The work further features several appendices providing supplementary details of the computations including the explicit expression for the six-loop anomalous dimensions.

\section{Brief review of the semiclassical approach} \label{review}

The aim of this section is to provide a general overview of the strategy we used to compute the anomalous dimension of fixed charge operators in the SM. In particular, we will briefly describe the semiclassical framework developed in \cite{Badel:2019oxl, Badel:2019khk, Antipin:2020abu, Antipin:2020rdw,  Antipin:2022naw, Antipin:2022hfe}.

Consider a quantum field theory invariant under the action of a certain global symmetry group $\mathcal{G}$ in $D$ dimensions. We collectively denote the couplings of the theory as $\kappa_I$. Any global symmetry implies the existence of a conserved current $\mathcal J^{\m}$ and a conserved charge $Q$ given by
\be\label{chargecons}
Q=\int d^{D-1} x\,\mathcal J^{0}(x) \,, \qquad \frac{d Q}{dt}=0 \,.
\ee
The method uses an expansion in one over the charge to determine quantities such as the anomalous dimension of certain fixed charge operators $\mathcal{O}_Q$.  To this extent, one exploits the fact that quantum systems can be well approximated by classical dynamics in the presence of large quantum numbers. 
In general, it is possible to fix up to rank$(\mathcal{G})$ charges, $Q_i$ which we can rescale as $Q_i = Q q_i$ in order to take $1/Q$ to be our small expansion parameter. The $\{q_i\}$ are a set of parameters of order unity that define the charge configuration and uniquely specify the irreducible representation according to which $\mathcal{O}_Q$ transforms. We first engineer a perturbative fixed point of the renormalization group (RG) flow where the theory is conformal and the anomalous dimensions are physical. We then make use of standard conformal field theory (CFT) tools to set up the $1/Q$ expansion and lastly, we rewrite our results for the anomalous dimensions as perturbative series in the running coupling constants. Crucially, in the latter form, the results remain valid outside the fixed point, i.e. for the real-world SM, and are shown to match the outcome of standard perturbation theory.

One starts by devising a perturbative fixed point of the  Wilson-Fisher type by tuning the mass parameters to $0$ and moving infinitesimally away from the upper critical dimension of the theory. In the SM, this amounts to setting the Higgs mass to $0$ and considering the theory in $D=4-\epsilon$ dimensions with $\epsilon \ll 1$. Generically the zeros of the beta functions are complex but, as has been shown in  \cite{Antipin:2020rdw}, we can still resort to the CFT toolbox. These allow us to determine the anomalous dimension of fixed-charge operators as a perturbative series in $\epsilon$. One can then use the fixed point values of the couplings $\kappa_I^*=\kappa_I^*(\epsilon)$  to express the results as a power series in the renormalized couplings $\kappa_I$.   The obtained results are valid away from the fixed point as confirmed diagrammatically to high-loop orders in e.g. \cite{Jack:2020wvs, Jack:2021lja, Bednyakov:2022guj, Antipin:2022hfe}.

We can now exploit the power of state-operator correspondence \cite{Cardy:1984rp} by first using Weyl invariance to map the theory onto a cylinder $\mathbb{R}\times S^{D-1}$. Concretely, if we use polar coordinates $(r, \Omega^{D-1})$ on $\mathbb{R}^D$ and we parameterize $\mathbb{R}\times S^{D-1}$ by  $(\tau, \Omega^{D-1})$, the map reads $r=R e^{\tau/R}$ with $R$ the radius of $S^{D-1}$. Note that Weyl invariance requires the introduction of a mass term for the scalar fields of the form $m^2 H^\dagger H$ where $m^2=(\frac{D-2}{2R})^2$. The latter arises from the coupling of the scalar fields to the Ricci scalar of $S^{D-1}$. For the fermions and gauge bosons, there are no extra correction terms when coupled to curved backgrounds and we just replace simple derivatives with the covariant ones.   Hence, the free Dirac action becomes $
S=\int_{\mathcal M} d^Dx\,\sqrt{-g}\,\,\bar\psi i\slashed\nabla_{\mathcal M}\psi \,,
$ where $\nabla_{\m}\psi\equiv\del_{\m}\psi+\frac{1}{4}\omega_{\m}^{ab}\g_{ab}\,\psi$ and $\slashed\nabla_{\mathcal M}\equiv \g^{\m}\nabla_{\m}$ is the Dirac operator on the manifold  $\mathcal M =\mathbb{R}\times S^{D-1}$ with $\g^{\m}$ \footnote{The infinitesimal Weyl transformations for the scalar, fermion, and the vierbein are given respectively by $\d\phi=\frac{D-2}{2}\s(x)\phi\,$, $\d\psi=\frac{(D-1)}{2}\s(x)\psi(x)$ and $\d e^a_{\m}=-\s(x)e^a_{\m}$. }.

Next, we use the state-operator correspondence \cite{Cardy:1984rp} to relate the scaling dimension $\Delta_{\mathcal{O}}$ of an operator $\mathcal{O}$ to the energy $E_\mathcal{O}$ of the corresponding state $\mathcal{O}(0) \ket{0}$ on the cylinder as $E_\mathcal{O}=\Delta_{\mathcal{O}}/R$. 
In particular, the scaling dimension $\Delta_{Q,\{q_i\}}$ of the lowest-lying operators carrying a given charge configuration $\left(Q, \{ q_i\}\right)$ is given by the ground state energy $E_{Q,\{q_i\}}$ of the theory on the cylinder in the presence of a finite charge density. We can obtain the latter by considering an arbitrary state $|{Q,\{q_i\}}\rangle$ with charges $Q_i=Q q_i$ and compute the expectation value of the evolution operator $e^{-HT }$ with $T$ the time interval and $H$ the Hamiltonian. It reads
\be\label{2point}
\langle Q,\{q_i\} |e^{-H T}|Q,\{q_i\}\rangle=\mathcal Z^{-1}\int  \mathcal D\Phi \,e^{-S_{\text{fixed}}[\Phi]} \,,
\ee
where $S_{\text{fixed}}$ is the charge fixed action, 
$\Phi$ represents collectively the field content of the theory, and $\mathcal{Z}$ is the partition function of the theory. Taking the  
$T \to \infty$ limit  the ground state saturates the above matrix element as
\be 
\label{ME}
\langle Q,\{q_i\} |e^{-H T}|Q,\{q_i\}\rangle\stackrel{T\to \infty}\sim e^{-E_{Q,\{q_i\}} T} = e^{-\frac{\Delta_{Q,\{q_i\}} T}{R}}  \,,
\ee
thereby yielding $\Delta_{Q,\{q_i\}}$. 

Having mapped the computation of the anomalous dimensions to the evaluation of the ground state energy of a finite density QFT on the cylinder, we now emphasize the advantages of the approach. The matrix element \eqref{ME} can now be computed via a semiclassical expansion equivalent to consider the following double-scaling limit \footnote{More precisely, since at the fixed point all the couplings are functions of $\epsilon$, the double scaling limit is \be 
Q\to\infty\,, \quad \epsilon \to 0 \,, \quad  Q \epsilon = (\text{fixed})\,.
\ee }
  
\be \label{dsl}
Q\to\infty\,, \quad \kappa_I \to 0 \,, \quad  Q \kappa_I= (\text{fixed})\,.
\ee 
Accordingly, the semiclassical expansion for the scaling dimension of the lowest-lying operator with charge $Q$ takes the form
\be
E_{Q,\{q_i\}} R=\Delta_{Q,\{q_i\}} = \sum_{j=-1} \frac{1}{Q^j}\Delta_j \left(Q\kappa_I^*, \{q_i\}\right) \, .
\label{ERexpansion}
\ee
The leading order $\D_{-1}$ is simply obtained by solving the equations of motion (EOMs) and plugging the solution into the action to get the classical energy. The subleading contribution $\D_0$ is given by the one-loop grand potential on the cylinder which is a Gaussian path integral over the fluctuations around the classical solutions. By varying the charge configuration $\{q_i\}$ one accesses different operators transforming according to different irreducible representations of $\mathcal{G}$. Noticeably, each $\Delta_{j}$ in eq.\eqref{ERexpansion} is the resummation of an infinite series of Feynman diagrams. Concretely, $\D_{-1}$ resums the terms with the leading power of $Q$ at every loop order, $\D_{0}$ resums the next-to-leading powers, and so on. In fact, the conventional perturbative series for the anomalous dimensions can be written as 
\be \label{PT}
\Delta_{Q,\{q_i\}} =Q\left(\frac{D-2}{2}\right) + \sum_{l=1} P_Q^{(l-\text{loop})} \qquad \text{where} \qquad P_Q^{(l-\text{loop})} = \sum_{k=0}^{l} C_{kl} Q^{l+1-k} \,.
\ee
By comparing to eq.\eqref{ERexpansion} one can see that the coefficients $C_{kl}$ appear in the small 't Hooft-like coupling $Q\kappa_I^*$ expansion of $\Delta_{k-1}$. At any loop order, there are $l+1$ coefficients $C_{kl}$ to compute. Two of them can be read off from the expressions for $\D_{-1}$ and $\D_{0}$. This opens the intriguing possibility of fixing the remaining $l-1$ unknown coefficients by matching eq.\eqref{PT} to the perturbative results for $l-1$ different values of $Q$ if they are known in the literature. 

In the opposite regime where $Q\kappa_I^*$ is taken to be large $\D_Q$ assumes the following general form
\begin{align}
\Delta_{Q}= Q^{\frac{D}{D-1}}\left[\alpha_{1}+ \alpha_{2} Q^{\frac{-2}{D-1}}+\alpha_3 Q^{\frac{-4}{D-1}}+\ldots\right] +Q^0\left[\beta_0+ \beta_{1} Q^{\frac{-2}{D-1}}+\ldots\right] + \mathcal{O}\left( Q^{-\frac{D}{D-1}}\right)   \,.
\label{largecharge}
\end{align}
The above structure can be also derived by considering an EFT description for the large charge sectors of generic interacting scalar CFT and is, therefore, insensitive to the microscopic dynamics. However, the latter is encapsulated in the parameters entering eq.\eqref{largecharge} which are related to the Wilson coefficient of the large charge EFT. At the same time, in odd dimensions, the coefficient $\beta_0$ is universal and can be predicted within the EFT framework \cite{Hellerman:2015nra}. Analogously, in even dimensions $\log(Q)$ terms with universal coefficients appear \cite{Cuomo:2020rgt}. The known field theories whose large charge sector is not described by the large charge EFT and do not satisfy eq.\eqref{largecharge} are non-interacting QFTs, supersymmetric theories where the lowest-lying state is related to a BPS operator, and certain fermionic models \cite{Dondi:2022zna}. 

Finally, in \cite{Antipin:2022hfe}, the approach has been generalized to $U(1)$ gauge theories in the concrete framework of the Abelian Higgs model in $D=4-\epsilon$ dimensions. In parallel, the construction of an on-shell conserved charge in Abelian gauge theories has been recently elucidated in \cite{Aoki:2022ugd}. There are two differences with respect to the Abelian global symmetry case. First, one needs to add a neutralizing charge background to avoid long-range electric fields producing infrared divergences. Being non-dynamical such a background only affects the EOM for the $U(1)$ gauge field as follows 
\be\label{neutralBG}
\del_\mu F^{\mu \nu} = J_{\text{tot}}^\nu =  J_{\text{Matter}}^\nu  -  J_{\text{Background}}^\nu = 0  \,,
\ee
where $F^{\mu \nu}$ and $ J^\nu$ are the electromagnetic field strength and current, respectively.
Secondly, while in the global symmetry case, $\Delta_Q$ corresponds to the scaling dimension of a certain local operator \footnote{For instance, in the $\lambda (\bar{\phi} \phi)^2$ theory at its weakly coupled Wilson-Fisher fixed point in $D=4-\epsilon$, $\Delta_Q$ is the conformal dimension of the $\phi^Q$ operator \cite{Badel:2019oxl}.}, when the $U(1)$ symmetry is gauged $\Delta_Q$ is related to a dressed two-point function \cite{Antipin:2022hfe}. For $Q=1$, the latter defines a non-local order parameter for the continuous phase transitions described by the model e.g. for the three-dimensional superconducting phase transition in the case of the Abelian Higgs model.

\section{Standard Model, notation and conventions} \label{StandardModel}

We are now ready to investigate the fixed hypercharge sector of the SM by starting with a concise review of the model Lagrangian. The latter is invariant under the gauge  $SU(3)_C\times SU(2)_L\times U(1)_Y$ symmetry and can be decomposed as follows
\be \label{STANDARD}
\mathcal{L}_{SM} = \mathcal{L}_{\text{kin}}   +\mathcal{L}_{H} +\mathcal{L}_{\text{Yukawa}} 
\,.
\ee
The gauge-fermion part of the Lagrangian is
\begin{align}
  \mathcal{L}_{\text{kin}} =& \ i\sum_{i=1,2,3} \left(   \bar{\mathcal{Q}}_i^L  \slashed D \mathcal{Q}_i^L +  \bar{u}_i^R \slashed D u_i^R +    \bar{d}_i^R \slashed D d_i^R +    \bar{L}_i^L  \slashed D L_i^L  +   \bar{l}_i^R \slashed D l_i^R \right) \nonumber \\ &  -\frac{1}{4}G_{\mu \nu}^A G^{\mu \nu \ A} -\frac{1}{4}B_{\mu \nu} B^{\mu \nu }-\frac{1}{4} W_{\mu \nu}^a W^{a\,\mu \nu } \,\,  \,,
\end{align}
where the covariant derivative is succinctly written as
\be
D_\mu =\del_\mu  
- i g^{'} \frac{Y}{2} B_\mu  - i {g}\frac{ \sigma^a}{2} W_\mu^a - i g_s T^A G_\mu^A \,\,.
\ee
Here $B_\m$, $W^a_\m$, and $G^A_\m$ are the gauge fields of  the $U(1)_Y$,  $SU(2)_L$ and  $SU(3)_C$ symmetries respectively with  $B_{\m\n}$, $ W^a_{\m\n}$ and $G^A_{\m\n}$ and  $(g^{'}, g\, , g_s)$ their corresponding gauge field strengths and  couplings. Also, $Y$ is the hypercharge, $\sigma^a$ are the Pauli matrices, and $T^A$ the Gell-Mann matrices. The $SU(2)_L$ doublets $\mathcal{Q}_i^L$ and $\mathcal{L}_i^L$ contain the left-handed quarks and leptons, respectively, while  $u_i^R$,  $d_i^R$,  $l_i^R$ are $SU(2)_L$ singlets corresponding to the right-handed SM fermions.  The  hypercharge values for the SM fields are gathered in the table below: \\

$\phantom{0000000000000}$\begin{tabularx}{0.6\textwidth} { 
  | >{\raggedright\arraybackslash}X 
  | >{\centering\arraybackslash}X 
  | >{\centering\arraybackslash}X 
  | >{\centering\arraybackslash}X 
  | >{\centering\arraybackslash}X  
  | >{\centering\arraybackslash}X 
  | >{\raggedleft\arraybackslash}X | }
 \hline
\text{Fields} & $H$ & $\mathcal{Q}^L$ & $u^R$ & $d^R$ & $L^L$ & $l^R$\\
 \hline
\,\,\,\,\,$Y$ & 1  & 1/3 & 4/3 & -2/3 & -1 & -2\,\,\,\\
\hline
\end{tabularx}\\

The Higgs-gauge sector of the SM is described by 
\begin{align}
    \mathcal{L}_{H}=(D_\mu H)^\dagger (D^\mu H )
- m^2 H^\dagger H- \frac{\lambda}{6}  \left( H ^\dagger H \right)^2 \ , 
\end{align}
with $H$ the $SU(2)_L$ doublet 
\begin{equation}
\label{Higgsdoublet}
    H= \left(\begin{array}{c}
    \phi_1 \\ \phi_2\end{array}\right) \ .
\end{equation}
Last but not least the Lagrangian describing Yukawa interactions reads
\be
\mathcal{L}_{\text{Yukawa}} = - \left(Y_u^{ij}\left(\bar{\mathcal{Q}}_i^L H^c \right)u_j^R + Y_d^{ij}\left(\bar{\mathcal{Q}}_i^L H \right)d_j^R+ Y_l^{ij}\left(\bar{L}_i^L H\right)l_j^R + \text{h.c.} \right)  \,,
\ee
where $H^c = i \sigma_2 H^\dagger$ is the charge-conjugated scalar field. The complex Yukawa matrices encode the flavor structure of the SM. All the couplings $ \kappa_I $ are renormalized as $ \kappa_{I0}  = M^\epsilon Z_{ \kappa_I } \kappa_I$ with $M$ the RG scale. We will consider a minimal subtraction renormalization scheme and make use of the one-loop value of the renormalization factors $Z_{ \kappa_I}$ or equivalently, the one-loop beta functions $\beta_{\kappa_I}$ which are summarized in Appendix \ref{appendicite}. Finally, we redefine the couplings as $\lambda \to (4 \pi)^2 \lambda$, $g \to 4 \pi g$,  $g^{\prime} \to 4 \pi g^\prime$,  $g_s \to 4 \pi g_s$, $Y_u \to 4 \pi Y_u $,  $Y_d \to 4 \pi Y_d $,  $Y_l \to 4 \pi Y_l $.
\section{Operator identification}
\label{OID}
The goal of this section is to identify the lowest-lying operator $\mathcal{O}_Q$ with hypercharge $Q$ whose anomalous dimension $\Delta_Q$ will be determined in this work. Since the precise form of the operator depends on which symmetries are gauged we start with the simple case of vanishing gauge couplings. Since the ground state we consider is spinless we are dealing with scalar operators. These are built out of tensor products of the Higgs fields decomposed according to irreducible representations  of the $SU(2)_L$ symmetry as follows: 
\be \label{product}
\mathbf{2}^{\otimes n} =\sum_{k=0}^{[n/2]}  
 a_k (\mathbf{n+1-2k} )  \,,
\ee
where the coefficients $a_k$ are the multiplicity of the representation in the tensor product decomposition. In the absence of the Yukawa couplings, the Higgs sector of the Standard Model is invariant under the symmetry group  $O(4)$. As shown in \cite{Antipin:2020abu}, the lowest-lying $O(4)$ operators with total charge $Q$ (defined as the sum of the individual $O(4)$ charges) have classical dimension $Q$ and are, therefore, built by multiplying the Higgs field with itself $Q$ times and do not contain derivatives. Moreover, they transform according to the traceless symmetric $O(4)$ representations with $Q$ indices. Then our operators appear in the branching of the latter in the $SU(2)_L \times U(1)_Y$ subalgebra of $O(4)$. In terms of the Dynkin labels for $O(4)$ ($D_2$) and $SU(2)_L$ ($A_1$), the branching rule reads
\be
[Q,0]_{D_2} \Rightarrow \sum_{k=0}^Q [Q-k]_{A_1}^{(Q-2k)}  \,,
\ee
where the superscript denotes the $U(1)_Y$ charge. Since the operators appearing on the right hand side have all classical dimension $Q$ by construction, the lowest-lying operators with $U(1)_Y$ charge $Q$ are obtained when $k=0$. We, therefore, conclude that the lowest-lying operators with $U(1)_Y$ charge $Q$ have no derivatives, classical scaling dimension Q, and transform according to the $SU(2)_L$ representations $[Q]_{A_1} = \mathbf{Q+1}$. Note that this corresponds to taking $n=Q$ and $k=0$ in eq.\eqref{product}. These operators can be compactly written as 
\be \label{tortellini}
\mathbf{Q+1} \Rightarrow \mathcal{O}_Q = H^{I_1} \dots  H^{I_Q}  \,.
\ee
For instance, the operator with $Q=1$ is the Higgs field itself, while the one with $Q=2$ transforms in the triplet (the adjoint) of $SU(2)_L$.

Now consider turning on the hypercharge gauge coupling $g^\prime$. As mentioned, the form of the relevant charged operator changes. In particular, in a CFT the $\Delta_Q$ characterizes the scaling of the two-point function of $\mathcal{O}_Q$ at criticality according to
\be \label{crit2pt}
G(x_i,x_f) = \braket{\mathcal{O}_Q(x_f) \mathcal{O}_Q(x_i)} = \frac{1}{\rvert x_f-x_i \rvert^{2 \Delta_Q}}  \,.
\ee
However, for $\mathcal{O}_Q$ given by eq.\eqref{tortellini}, the above correlator is gauge-dependent and evaluates to zero due to Elitzur's theorem \cite{Elitzur:1975im}. It is therefore necessary to dress the two-point function via the introduction of a gauge line connecting the two external points as
\be \label{dresso}
G(x_i,x_f) =\braket{ H^{I_1} \dots  H^{I_Q}(x_f) \exp\left(i \ g^\prime\,Q\int_{x_i}^{x_f} d^D x \,J^\mu (x) B_\mu(x)\right) H^{I_1} \dots  H^{I_Q}(x_i)} \,,
\ee
where $J^\mu$ satisfies the  Ward identity
\be
\partial_\mu J^\mu = \delta(x-x_f) - \delta(x-x_i)  \,,
\ee
which ensures the gauge invariance of the two-point function. Therefore, the problem of determining the lowest-lying operator with hypercharge $Q$ turned into the problem of determining $J^\mu$ such that $G(x_i,x_f)$ scales according to the smallest conformal dimension. 
It has been shown in \cite{Antipin:2022hfe} that the smallest scaling dimension is achieved by considering a \emph{Dirac line} \cite{Dirac}. The latter satisfies the $\del^2J_{\m}=0$ condition while the explicit form of the non-local current is $J_\mu = J^{'}_{\m}(z- x_f) - J^{'}_{\m}(z- x_i)$ where
\be
J^{'}_{\m}(z) =-i \int \frac{d^D k}{(2 \pi)^D} \frac{k_\mu}{k^2} e^{i k \cdot z}=- \frac{\Gamma(D/2-1)}{4 \pi^{D/2}}\del_\mu \frac{1}{z^{D-2}}\,\,.
\ee
 
This construction allows us to identify $\Delta_Q$ as the anomalous dimension of the dressed operator
\be\label{order}
\mathcal{O}_Q(x) =  e^{-i g^\prime Q \int d^Dz J^{'}_{\m}(z-x)B^{\mu}(z)} H^{I_1} \dots  H^{I_Q}  \,,
\ee
whose two-point function at criticality is given by eq.\eqref{crit2pt}. The above operator can be interpreted as the insertion of $Q$ units of charge at the point $x$ dressed with a coherent state of photons describing the induced Coulomb field. Interestingly, since $J^{'}_{\m}$ is a total derivative, the two-point function \eqref{dresso} reduces to the gauge-dependent correlator $\braket{ H^{I_1} \dots  H^{I_Q}(x_f) H^{I_1} \dots  H^{I_Q}(x_i)}$ in the Landau gauge $\partial^\mu B_\mu = 0$. As a consequence, the perturbative results for $\Delta_Q$ match the scaling dimension of the local operator \eqref{tortellini} in the Landau gauge.

Since the operator \eqref{order} transforms nontrivially under $SU(2)_L$, when the latter symmetry is gauged due to a non-vanishing value of $g$, $\Delta_Q$ no longer corresponds to its conformal dimension. As a consequence, although the computations are still valid, the identification of the lowest-lying operator with hypercharge $Q$ becomes a subtle issue that goes beyond the scope of this work \footnote{For instance, we note that the construction of a non-Abelian counterpart of the Abelian Dirac line, which has been suggested as the order parameter for the confining phase of QCD, is an interesting open problem \cite{Caudy:2007sf, Greensite:2017ajx}.}. In what follows, we will, however, make use of the $g \to 0$ limit, for which we can identify the relevant two-point function as eq.\eqref{dresso}, to partly check the validity of our calculations against diagrammatic calculations. Concretely, for $g=0$ and in the Landau gauge we will show that our results for $\Delta_Q$ match the three-loop anomalous dimension of the operators defined in \eqref{tortellini} \cite{sasha}. Moreover, the $Q=1$ case yields the anomalous dimension of the Higgs field which has been computed to high precision in e.g. \cite{Bednyakov:2013eba, Chetyrkin:2012rz, Bednyakov:2013cpa}. Finally, we stress that no operator identification issue arises when the weak symmetry is global providing relevant results for testing higher-order computations stemming from  $\lambda$, $g_S$, and $y_t$.

\section{Scaling dimensions at leading order} \label{LO}

Our goal is to compute the scaling dimension $\Delta_Q$ of the lowest-lying operator with a certain value of the hypercharge $Q$. According to the discussion in Sec.~\ref{review}, in the double scaling limit \eqref{dsl} the scaling dimension takes the form \eqref{ERexpansion} with the leading term given by the classical ground state energy of the conformal theory on $\mathbb{R}\times S^{D-1}$. In what follows, we will measure all the dimensionful quantities in units of the radius of $S^{D-1}$ which we set equal to unity. As previously discussed, conformal symmetry on the cylinder is achieved by tuning the Higgs mass term to $m^2=\left(\frac{d-2}{2}\right)^2$ and engineering a complex zero for all the SM beta functions (which we summarize in App.\ref{appendicite}) in $D=4-\epsilon$ dimensions. We stress again that this procedure is just a computational trick and the final results will apply to the real-world SM as well. The hypercharge is fixed by introducing into the SM Lagrangian \eqref{STANDARD} the associated chemical potential as the temporal component of a background $U(1)_Y$ gauge field, i.e. by upgrading the covariant derivative as
\be
D_\mu \to D_\mu - i \mu Y  \,.
\ee
The relevant equations of motion are
\begin{align}
  \partial_\nu B^{\mu \nu} &= 0  \ , \\ 
  -(D_\mu-i\mu\delta_{\mu0})(D^\mu-i\mu\delta^{\mu0})H-m^2 H-\frac{(4 \pi )^2 }{3}  \lambda
(H^\dagger H)H &=0 \ , \\
\partial^\mu W_{\mu\nu}^{(a)}+4\pi g\epsilon^{abc}A^{\mu(b)} W_{\mu\nu}^{(c)}
+4 \pi i g\left[H^\dagger\frac{\tau^a}{2}\partial_\nu H-\partial_\nu
H^\dagger\frac{\tau^a}{2}H\right] & \nonumber \\ +\frac{g^2 (4 \pi )^2}{2}W_{\nu}^{(a)}
H^\dagger H
+8 \pi g\mu\delta_{\nu0}H^\dagger\frac{\tau^a}{2}H &=0  \ .
\end{align}

As discussed in Sec.~\ref{review}, we enforced charge neutrality by introducing a background current which makes the RHS of the first EOM vanishing. Moreover, we assumed a zero classical value for both the gluon field (since it does not couple directly to the Higgs) and the fermions. This trivializes all the EOMs except the above. As shown in \cite{Gusynin:2003yu}, the global minimum of the action is achieved by a vacuum solution that breaks rotational invariance and is described by the following ansatz
\begin{align}
& W_3^{(+)}=(W_3^{(-)})^*=C\neq0 \ ,\quad W_0^{(3)}=P\neq0 \ ,
\quad H=\frac{1}{\sqrt{2}}\left(\begin{array}{c}
   0 \\ v\end{array} \right) \ , \nonumber \\
&  v \neq 0 \ , \quad  W_{1,2}^{(\pm)}
=W_{0}^{(\pm)}=W_{1,2}^{(3)}=W_{3}^{(3)}=0 \ , \quad B_\mu = 0   \,,
\label{ansatz}     
\end{align}
where we introduced $W_\mu^{(\mp)}=\frac{1}{\sqrt{2}}(W_\mu^{(1)}\pm i W_\mu^{(2)})$. The various vacuum expectation values are determined by the EOMs as
\begin{align}
\left(P^2-\frac{v^2}{4}\right)C &=0 \ ,\\
\left(2|C|^2+\frac{v^2}{4}\right)P-\frac{ \mu v^2}{8 \pi g} 
 &= 0 \ ,\\
\left[\left(\mu-2 \pi g P\right)^{2} -m^2-\frac{(4 \pi )^2}{6} \lambda v^2-\frac{(4 \pi )^2}{2} g^2 |C|^2 \right]v &=0  \,,
\label{anzats-eq}    
\end{align}
with solution
\be
 P=\frac{v}{2} \ ,\quad |C|^2 = v \left(\frac{\mu }{8 \pi g}-\frac{v}{8}\right)  \ ,\quad  \tilde{v} \equiv 4\pi v= \frac{9 g \mu -\sqrt{96  \lambda  \left(\mu ^2-m^2\right)+9 g^2 \left(\mu ^2+8 m^2\right)}}{3 g^2-4 \lambda }  \,.
\label{solution-for-A03}
\ee
The classical action evaluated on the above solution reads
\begin{equation} \label{S0}
   \mathcal{S}^{(0)} =   \frac{ m^2}{16} \tilde{v}^2 -\frac{1}{16}  \left(\frac{g \tilde{v}}{4}- \mu \right)^2  \tilde{v}^2+ \frac{\lambda}{192}   \tilde{v}^4   + \mu  Q  \,.
\end{equation}
Finally, the relation between $Q$, $\mu$, and $v$ is obtained via the last EOM $\partial  \mathcal{S}^{(0)} / \partial \mu = 0$, which yields
\be
\mu  \tilde{v}^2 - \frac{g \tilde{v}^3}{4}=8 Q  \,.
\ee

One can easily check that in the limit $g \to 0$ vector condensation does not occur, i.e. $C=P=0$. In this case the ground state is isotropic, homogeneous, and characterized by scalar condensation with the vev of the scalar field $v$, and the classical action $\mathcal{S}$ given by eq.\eqref{solution-for-A03} and eq.\eqref{S0} with $g=0$. The solution with $g=0$ and the associated symmetry-breaking pattern define a phase of matter dubbed \emph{conformal superfluid phase} describing, along with its non-Abelian generalization, the large-charge sector of most CFTs considered in the literature \cite{Hellerman:2015nra,  Monin:2016jmo, Badel:2019khk, Jack:2021lja, Antipin:2020rdw, Badel:2019oxl, Antipin:2022naw, Antipin:2022hfe, Antipin:2020abu}. On the other hand, the large-charge sector of the Weinberg-Salam model realizes a different phase characterized by vector condensation and the associated breaking of spatial rotations \cite{Sannino:2002wp,Gusynin:2003yu}. Indeed, the solution \eqref{ansatz} is spatially homogeneous but anisotropic and breaks $SO(3) \times SU(2)_L \times U(1)_Y$ down to $SO(2)$ (spatial rotations in the $x_1-x_2$ plane). The spectrum of excitations has been studied in \cite{Sannino:2002wp} for massive vectors fields and in \cite{Gusynin:2003yu} for the SM. It contains two massless relativistic modes. In other words, the Goldstone theorem works as usual when $SO(3)$ is viewed as a global symmetry\footnote{For a general counting of the Goldstone bosons versus different  phases that relativistic vector condensation can feature see \cite{Sannino:2002wp} }.

A few comments are in order. 
\begin{itemize}

\item The phase we consider is not the phase usually discussed in the context of SM thermodynamics \cite{Loewe:2004zw, Kapusta:1990qc, Kapusta:1981aa}. In fact, when the Higgs mass term is negative, the SM is in the broken phase already at $\mu=0$ and it is possible to realize a phase with unbroken rotational invariance. The corresponding classical solution reads
\be
\tilde{v}^2=-\frac{6m^2}{\lambda}  \,, \quad
W_0^{(3)}=\frac{\mu}{2 \pi g}  \,,\quad W_0^{(\pm)}=0 \,, \quad B_\mu =0   \,.
\label{vacuum2}
\ee
This solution provides the ground state of the SM for low values of the chemical potential $\mu^2 < \frac{3g^2}{8 \l}\rvert m^2 \rvert$. However, at the critical point $\mu^2 = \frac{3g^2}{8  \l}\rvert m^2 \rvert$ a second order phase transition to a phase with broken rotational invariance occurs \cite{Gusynin:2003yu}.

\item The vev for the Higgs field \eqref{solution-for-A03} diverges at $g^2= \frac{4}{3}  \lambda$ signaling a phase transition. While for $g^2 > \frac{4}{3} \lambda$, the solution \eqref{solution-for-A03} is still a local minimum of the potential, the latter becomes unbounded from below. In what follows we will therefore assume $g^2 < \frac{4}{3} \lambda$. However, this restriction will not affect our results for the anomalous dimensions which hold for arbitrary values of the coupling.

\item One may try to avoid spontaneous breaking of rotational invariance by enforcing non-Abelian charge neutrality via the introduction of a non-Abelian background current $J_\mu^a$. However, this leads to inconsistencies in the spectrum of physical modes  \cite{Hama:2011rt}. In fact, as pointed out in \cite{Watanabe:2014qla}, in the non-Abelian case, it is not possible to include a non-dynamical charge background without violating the symmetry of the theory.

\end{itemize}

Let us recap the calculation of $\Delta_{-1}$ in the $g \to 0$ limit where the solution to the EOMs is
\be \label{g0EOM}
\mu= \frac{y^{2/3}+3^{1/3}}{3^{2/3} y^{1/3}}  \ , \quad y=6 \l Q + \sqrt{(6 \l Q)^2 -3}  \,,
\ee
whereas the classical action reads 
\be \label{energz}
\mathcal{S}^{(0)} = \frac{Q \left(3 \mu ^2+m^2\right)}{4 \mu }  \,.
\ee
Then the leading order in the semiclassical expansion is obtained by substituting the solution of eq.\eqref{g0EOM} into the above yielding
\be
 \label{classic}
 Q\,\Delta_{-1}=  \frac{6 \ 3^{1/3} y^{2/3}-3 y^{4/3}+3^{1/3} y^{8/3}+2\ 3^{2/3} y^2+3\ 3^{2/3}}{48 y^{4/3}  \lambda }  \ .
\ee
This result holds in many models with $O(N)$ symmetry in $D=4-\epsilon$ dimensions \cite{Badel:2019oxl, Antipin:2020abu, Antipin:2022hfe, Antipin:2022naw, Jack:2021lja} where it resums the terms with the highest power of $Q$ to all orders in perturbation theory. For the $g \neq 0$ case, we obtain
\begin{align} \label{LEAD}
 Q  \Delta_{-1} &= \frac{3}{32 \left(3 g^2-4 \lambda \right)^3} \Bigg(9 g^4 \left(3 \mu_0 ^4-20 \mu_0 ^2+8\right)-48 g^2 \lambda  \left(15 \mu_0 ^4-\mu_0 ^2+4\right) \nonumber \\ &+128 \lambda ^2 \left(-3 \mu_0^4+2 \mu_0^2+1\right)+ 3 g \mu_0 ^3 \left(3 g^2 + 32 \lambda \right) \sqrt{9 g^2 \left(\mu_0 ^2+8\right)+96 \lambda  \left(\mu_0^2-1\right)}\Bigg)  \,,
\end{align}
where $\mu_0$ is the real solution of 
\be \label{theequation}
\frac{\mu_0  \left(9 g \mu_0  -\sqrt{9 g^2 \left(\mu_0  ^2+8\right)+96 \lambda  \left(\mu_0  ^2-1\right)}\right)^2}{8 \left(3 g^2-4 \lambda \right)^2}-\frac{g \left(9 g \mu_0 -\sqrt{9 g^2 \left(\mu_0  ^2+8\right)+96 \lambda  \left(\mu_0 ^2-1\right)}\right)^3}{32 \left(3 g^2-4 \lambda \right)^3}=Q  \,,
\ee
which minimizes $\Delta_{-1}$. Unlike the $g= 0$ case, it is not possible to obtain a  compact expression for $\Delta_{-1}$ as a function of the 't Hooft-like couplings $\lambda Q$ and $g \sqrt{Q}$. However, it is possible to solve eq.\eqref{theequation} perturbatively for small values of the 't Hooft-like couplings and determine $\Delta_{-1}$ to arbitrarily high orders in such expansion. As discussed in Sec.~\ref{review}, this is the regime where we recover ordinary diagrammatic results. In Sec.~\ref{disco} we will provide an explicit expression for the small 't Hooft-like coupling expansion of $\Delta_{-1}$ up to $3$ loops and check it against perturbation theory. Novel additional results up to the six-loop order are collected in App.\ref{C}. 

\section{Scaling dimensions at next-to-leading order} \label{NLO}
The next-to-leading order $\Delta_0$ consists of the Gaussian path integral over the fluctuations on top of the classical solution. Concretely, it can be written as the sum of the zero-point energies.  For this involved contribution the $SU(2)_L$ gauge coupling $g$ is set to zero. This allows us to provide a more transparent presentation of the key issues related to this computation while setting the stage for the complete result in a forthcoming work. 

Since bosons and fermions do not mix at the quadratic level, we can separate their contributions as
\begin{align}
 \Delta_0  =  \Delta_0^{(bos)} -  \Delta_0^{(fm)} \,,
 \end{align}
 where \cite{Badel:2019oxl, Antipin:2020abu}
 \begin{align} \label{general0}
 \Delta_0^{(fm)} = \frac12 \sum_{i=\text{fermions}}  \sum_{\ell=\ell_0}^\infty n_{i}(\ell) \omega_i(\ell) \,, \qquad \Delta_0^{(bos)} =\frac12  \sum_{i=\text{bosons}} \sum_{\ell=\ell_0}^\infty n_{i}(\ell) \omega_i(\ell) \,,
\end{align}
where the index $i$ runs over all the nontrivial dispersion relations ($\omega \neq p$ with $p$ the spatial momentum) $\omega_i$ of the spectrum. Here $\ell$ labels the eigenvalues of the relevant differential operator (Laplacian and Dirac operators for bosons and fermions, respectively) on $S^{D-1}$ which have multiplicity $n_i(\ell)$. The relevant expressions for fields of different spin are collected in App.\ref{sphere}. In the next section, we illustrate the calculation of $\Delta_0^{(fm)}$ in the case of a single generation of fermions.

\subsection{Warm-up: single generation Yukawa contribution}

The Yukawa interactions for the third generation of fermions read
\begin{align}
\mathcal{L}_{\text{Yukawa}} =&-4 \pi y_t \bar {\mathcal{Q}}^L H^c t_R - 4 \pi y_b \bar{ \mathcal{Q}}^L H b_R- 4 \pi y_\tau \bar L^L H \tau_R + \text{h.c.} \nonumber \\ =& - 4 \pi y_t \left[(\bar t P_R t) \phi_2^* +(\bar t P_L t) \phi_2 -(\bar b P_R t) \phi_1^* - (\bar t P_L b) \phi_1 \right]\nonumber\\ 
&- 4 \pi y_b \left[(\bar b P_R b) \phi_2+(\bar b P_L b) \phi_2^* -(\bar t P_R b) \phi_1 - (\bar b P_L t) \phi_1^* \right]\nonumber \\ 
&-  4 \pi y_\tau \left[(\bar \tau P_R \tau) \phi_2 +(\bar \tau P_L \tau) \phi_2^* -(\bar \nu_{\tau} P_R \tau) \phi_1 - (\bar \tau P_L \nu_\tau) \phi_1^* \right] \ .
\end{align}
 Here $\mathcal{Q}^L=(t,b)_L$ and $L^L=(\nu_\tau, \tau)_L$ are the $SU(2)_L$ left-handed doublets and $P_{L/R} = \frac{1}{2}\left(1 \pm \gamma_5 \right)$. For the quadratic Lagrangian of the fermions we have
 \begin{align}
\mathcal{L}_{fm}^{(2)} &=i \bar{t} \slashed \nabla t- \mu Y_{\mathcal{Q}_L} \bar{t} \gamma_0 P_L t - \mu Y_{u_r} \bar{t} \gamma_0 P_R t + \frac{y_t }{\sqrt{2}} \tilde{v}  \bar{t} t   + i \bar{b} \slashed \nabla b - \mu Y_{\mathcal{Q}_L} \bar{b} \gamma_0 P_L b - \mu Y_{d_r} \bar{b} \gamma_0 P_R b  \nonumber \\ &+ \frac{y_b}{\sqrt{2}} \tilde{v}  \bar{b} b     + i \bar{\tau} \slashed \nabla \tau - \mu Y_{L_L} \bar{\tau} \gamma_0 P_L \tau - \mu Y_{l_r} \bar{\tau} \gamma_0 P_R \tau + \frac{y_\tau }{\sqrt{2}} \tilde{v}  \bar{\tau} \tau   +i \bar{\nu_\tau} \slashed \nabla \nu_\tau    - \mu Y_{L_L} \bar{\nu_\tau} \gamma_0 P_L \nu_\tau   \,,
 \end{align}
where the scalar vev is given by eq.\eqref{solution-for-A03} for $g=0$, that is
\be
\tilde{v}^2 = \frac{6}{\lambda } \left(\mu ^2-m^2\right) \,.
\ee
It is possible to redefine the left-handed neutrino field as $\nu_\tau \to e^{- Y_{L_L} \mu t} \nu_\tau$, $\bar{\nu}_\tau \to e^{ Y_{L_L} \mu t} \bar{\nu}_\tau$ to eliminate the last term from the Lagrangian. Then the neutrino has a trivial dispersion relation $\omega=p$ and does not contribute to $\Delta_0$. The remaining fermions do not mix among themselves and have the same quadratic Lagrangian except for the hypercharge assignation and the relevant Yukawa coupling. The quadratic Lagrangian for the top quark field entails the following inverse propagator
\be
D^{-1} = \slashed p  - \mu Y_{\mathcal{Q}_L} \gamma_0 P_L - \mu Y_{u_R} \gamma_0 P_R + \frac{\tilde{v}}{\sqrt{2}} y_t \mathbbm{1}_{4 \times 4}  \,.
\ee
The dispersion relations are found by solving $\det D_i^{-1}= 0$ for every fermion. Noticeably, despite the different values of the hypercharge, the dispersion relations have all the same form up to an irrelevant term that does not contribute to $\Delta_Q$. After dropping the latter, we arrive at the following dispersion relations
\be \label{confradioferm}
\omega_{\pm}^{(i)}(\ell) = \sqrt{\left(\frac{\mu }{2}+\l_{f\pm}\right)^2+ \frac{y_i^2 \tilde{v}^2}{2}}\,\,.
\ee
where $i=t, b, \tau$ and $\l_{f\pm}$ denotes the eigenvalues of the Dirac operator on $S^{D-1}$ whose value and degeneracy are given in eq.\eqref{eigenforferm}.
Since all the dispersion relations have the same form the Yukawa contributions to $\Delta_0$ can be divided as 
\begin{align} \label{square}
\Delta_0^{(fm)} = N \Delta_0^{(t)} + N \Delta_0^{(b)}  +  \Delta_0^{(\tau)}\,,
\end{align}
where $N$ stands for the number of colors and
\be
\Delta_0^{(t)} =\frac12\sum_{\ell=0} n_f(\ell) \left(\omega_{+}^{(t)} (\ell)+ \omega_{-}^{(t)}(\ell)\right)\,, \qquad \Delta_0^{(b)} = \Delta_0^{(t)}(y_t \to y_b)  \,, \qquad \Delta_0^{(\tau)} = \Delta_0^{(t)}(y_t \to y_\tau)  \,.
\ee
The sum over $\ell$ needs to be regularized and renormalized. The former step can be easily performed by subtracting the divergent terms in the expansion of the summand around $\ell = \infty$. The sum over subtracted terms is then zeta-regularized and added back to the result. However, the "logarithmic divergence" $\sim \sum_{\ell = 1} \frac{1}{\ell}$ needs to be further regularized using dimensional regularization as $\sum_{\ell = 1}\frac{1}{\ell} =  \zeta(1 +\epsilon) = \frac{1}{\epsilon} + \gamma_E$ with $\gamma_E$ the Euler-Mascheroni constant. This allows isolating a divergent term scaling as $\frac{1}{\epsilon}$ which is canceled when renormalizing the classical contribution $\Delta_{-1}$. We obtain
\begin{align}
    \Delta_0^{(t)} &=\sqrt{\frac{12 \left(\mu ^2-1\right) y_t^2}{\l}+(\mu -3)^2}+\sqrt{\frac{12 \left(\mu ^2-1\right) y_t^2}{\l}+(\mu +3)^2} \nonumber \\ & -\frac{3 \left(\mu ^2-1\right) y_t^2 \left(\l \left(13-3 \mu ^2\right)+3 \left(3 \mu ^2+1\right) y_t^2\right)}{8 \l^2}-6 +\frac{1}{2} \sum_{\ell=1}^\infty\sigma^{(t)} \,,
\end{align}
where
\be
\begin{split}
\sigma^{(t)}(\ell) 
=&  \frac{1}{2 \l^2 \ell}\Big(-4 \l^2 \ell (\ell+1) (\ell+2) (2 \ell+3)-3 \l \left(\mu ^2-1\right) y_t^2 \left(4 \ell^2+\mu ^2+6 \ell-1\right)+9 \left(\mu ^2-1\right)^2 y_t^4\Big)\\
&+2 (\ell+1) (\ell+2) \left(\sqrt{\frac{3 \left(\mu ^2-1\right) y_t^2}{\l}+\left(\frac{2 \ell+3-\mu}{2} \right)^2}+\sqrt{\frac{3 \left(\mu ^2-1\right) y_t^2}{\l}+\left(\frac{2 \ell+3+\mu}{2} \right)^2} \right) \,.
\end{split}
\ee

 Having solved the one-generation case we can now tackle the full flavour sector of the SM. 
\subsection{Complex Yukawa matrices} 
The flavour structure of the SM  is encoded in the Yukawa couplings summarised as three complex $n_G \times n_G$ matrices where $n_G$ is the number of generations yielding the following Lagrangian
\be
\mathcal{L}_{\text{Yukawa}} = - 4 \pi \left(Y_u^{ij}\left(\mathcal{Q}_i^L H^c \right)u_j^R + Y_d^{ij}\left(\mathcal{Q}_i^L H \right)d_j^R+ Y_l^{ij}\left(L_i^L H\right)l_j^R \right)   \,.
\ee
As we shall see, our results will depend on the following traces of Yukawa matrices
\be
\mathcal{Y}_{f} = (4\pi)^2\Tr Y_f Y_f^\dagger \,, \quad \mathcal{Y}_{ff} =  (4\pi)^4\Tr (Y_f Y_f^\dagger)^2    \,, \quad \mathcal{Y}_{fff} =  (4\pi)^6\Tr (Y_f Y_f^\dagger)^3 \,, \quad f = u, d, l \,.
\ee
Analogously to the $n_G=1$ case, the fermionic contribution can be divided as
\begin{align} \label{ineq}
\Delta_0^{(fm)} &= N \left( \Delta_0^{(f=u)} + \Delta_0^{(f=d)} \right)  +\Delta_0^{(f=l)}\,, \\ \Delta_0^{(f)}  &=\frac12\sum_{i=1}^{n_G} \sum_{\ell = 0} n_f(\ell) \left(\omega^{(i)}_{f+}(\ell) + \omega^{(i)}_{f-}(\ell) \right )  \,.
\end{align}
For each value of $f$ the inverse propagator is a $4n_G \times 4 n_G$ matrix where the off-diagonal $4\times 4$ blocks contain the mixing among different generations owed to the Yukawa structure. For instance, the inverse propagator for the up quarks in the $n_G=3$ case reads
\begin{align}
    D^{-1} = \left(\begin{array}{ccc}
      U_1   & V_{12} & V_{13}  \\
      V_{21}   & U_2  & V_{23}\\
        V_{31}   & V_{32} & U_3   
    \end{array} \right)  \,,
\end{align}
where
\begin{align}
     U_i  &=  \slashed p  - \mu Y_{\mathcal{Q}_L} \gamma_0 P_L - \mu Y_{u_R} \gamma_0 P_R + \frac{\tilde{v}}{\sqrt{2}} Y_u^{ii} P_L + \frac{\tilde{v}}{\sqrt{2}} \left(Y_u^{ii}\right)^* P_R \,,  \qquad i = 1,2,3  \\ 
     V_{ij} &= \frac{\tilde{v}}{\sqrt{2}} \text{diag}\left( Y_u^{ij},  Y_u^{ij}, \left(Y_u^{ji}\right)^*, \left(Y_u^{ji}\right)^*  \right)  \,.
\end{align}
For the two-generation case, the dispersion relations read
\begin{align}
  \omega_{f\pm}^{(1)}(\ell) &=  \sqrt{\left( \frac{\mu}{2}+\lambda_{f \pm} \right)^2 +\frac{ \tilde{v}^2}{4} \left(\mathcal{Y}_f+\sqrt{2 \mathcal{Y}_{ff}-\mathcal{Y}_f^2}\right)}   \,,\\
 \omega_{f\pm}^{(1)}(\ell) &=  \sqrt{\left( \frac{\mu}{2}+\lambda_{f \pm} \right)^2 +\frac{\tilde{v}^2 }{4} \left(\mathcal{Y}_f-\sqrt{2 \mathcal{Y}_{ff}-\mathcal{Y}_f^2}\right)}  \,,
\end{align}
whereas in the complete $n_G=3$ case we obtain
{\small{
\begin{align}
  \omega_{f\pm}^{(1)}(\ell) &=\sqrt{\left( \frac{\mu}{2}+\lambda_{f \pm} \right)^2 -\frac{\tilde{v}^2}{3A} \left(A^2 - 2 A \mathcal{Y}_f - 2 \mathcal{Y}_f^2 + 6 \mathcal{Y}_{ff}\right)}  \,, \\
  \omega_{f\pm}^{(2)}(\ell) &=\sqrt{\left( \frac{\mu}{2}+\lambda_{f \pm} \right)^2 + \frac{\tilde{v}^2}{24 A} \left(4 A \mathcal{Y}_f+\left(1-i \sqrt{3}\right) A^2 -2\left(1+i \sqrt{3}\right) \left(\mathcal{Y}_f ^2-3 \mathcal{Y}_{ff} \right)\right)}  \,, \\
  \omega_{f\pm}^{(3)}(\ell) &=\sqrt{\left( \frac{\mu}{2}+\lambda_{f \pm} \right)^2 + \frac{\tilde{v}^2}{24 A} \left(4 A \mathcal{Y}_f+\left(1+i \sqrt{3}\right) A^2 -2\left(1-i \sqrt{3}\right) \left(\mathcal{Y}_f ^2-3 \mathcal{Y}_{ff} \right)\right)}  \,,
\end{align}}} \normalsize

where

\begin{align}
A =&\bigg(6 \sqrt{2} \sqrt{\mathcal{Y}_f^6-9 \mathcal{Y}_f^4 \mathcal{Y}_{ff}+8 \mathcal{Y}_f^3 \mathcal{Y}_{fff}+21 \mathcal{Y}_f^2 \mathcal{Y}_{ff}^2-36 \mathcal{Y}_f \mathcal{Y}_{ff} \mathcal{Y}_{fff}-3 \mathcal{Y}_{ff}^3+18 \mathcal{Y}_{fff}^2}\nonumber \\ & +36 \mathcal{Y}_f \mathcal{Y}_{ff}-36 \mathcal{Y}_{fff} -8 \mathcal{Y}_f^3\bigg)^{1/3}  \,.
\end{align}
By regularizing the sum over $\ell$ in eq.\eqref{ineq} and renormalizing the resulting expression as explained in the previous section we can write $\Delta_0^{(f)}$ in terms of a convergent sum as
\begin{align} \label{NLOFERM}
    \Delta_0^{(f)} &=  \frac{9 \left(-3 \mu ^4+2 \mu ^2+1\right)\mathcal{Y}_{ff} }{8 \l^2}+\frac{3 \left(3 \mu ^4-16 \mu ^2+13\right) \mathcal{Y}_f}{8 \l}-18 \nonumber \\ &+ \frac12\sum_{i=1}^{3} n_f(0) \left(\omega_{f+}^{(i)}(0) + \omega_{f-}^{(i)} (0) \right)+ \frac{1}{2} \sum_{\ell=1} \sigma(\ell) \,,
\end{align}
where
\begin{align} \label{SUMFERM}
     \sigma(\ell) &= -6 (\ell+1) (\ell+2) (2 \ell+3)-\frac{\mu ^2-1}{2 \l^2 \ell} \left( 3 \l \mathcal{Y}_{f} \left(4 \ell^2+6 \ell+\mu ^2-1\right)+9 \left(\mu ^2-1\right) \mathcal{Y}_{ff} \right)\nonumber \\ &+ 2 (\ell+1) (\ell+2) \sum_{i=1}^{3} \left[\omega_{f+}^{(i)}(\ell) + \omega_{f-}^{(i)} (\ell)  \right]_{d=4} \,.
\end{align}

\subsection{Scalars and vector bosons contributions}
\label{gauged hypercharge}

In this section, we present the contributions to the anomalous dimensions from the bosonic sector of the theory by expanding around the classical solutions as
\be\label{phi1phi}
\phi_1=\frac{1}{\sqrt2}e^{\m\tau}(\varphi_1(x)+i\varphi_2(x))\,,\qq \phi_1^{*}=\frac{1}{\sqrt2}e^{-\m\tau}(\varphi_1(x)-i\varphi_2(x)) \,,
\ee
and
\be\label{phi2rpi}
\phi_2=\frac{1}{\sqrt2}(v+r(x))e^{\m\tau+i\p(x)/v} \,, \qq\phi_2^{*}=\frac{1}{\sqrt2}(v+r(x))e^{-\m\tau-i\p(x)/v} \,,
\ee
where $r,\p$ and $\varphi_{(1,2)}$ are real valued fields. Using  the expressions \eqref{phi1phi} and \eqref{phi2rpi}, the quadratic part of the Lagrangian is 
\be\label{LquadraticU(1)}
\mathcal L^{(2)}\left
(\varphi,r,\p, B_\m\right)=\mathcal L^{(2)}(\varphi_1,\varphi_2)+\mathcal L^{(2)}(r,\p, B_\m) \,,
\ee
where
\be
\mathcal L^{(2)}
(\varphi_1,\varphi_2)=\frac{1}{2}\varphi_1(-\nabla^2)\varphi_1+\frac{1}{2}\varphi_2(-\nabla^2)\varphi_2
+\frac{1}{2}2i\m\varphi_2\dot\varphi_1-\frac{1}{2}2i\m\varphi^1\dot\varphi^2 \,,
\ee
and 
\be\begin{split}
\mathcal L^{(2)}(r,\p,B_\m)&=\frac{1}{4}B_{\m\n}^2+\frac{1}{2}r(-\nabla^2) r+\frac{1}{2}\p(-\nabla^2)\p+i\m \,g{'}\tilde vYB_{0}r\\
&-\frac12g{'}\tilde vYB_{\m}\nabla^{\m}\p
-\frac{i}{2}4\,\m\, r\dot\p-(m^2-\m^2)r^2+\frac{1}{8}(g{'}\tilde vY)^2 B_{\m}^2\,\,.
\end{split}
\ee
Note that at the quadratic level the $\varphi_{(1,2)}$ decouple from the rest of the action. Therefore, our results simplify as follows
\be
\mathcal Z_{\text{1-loop
}}=\mathcal Z^\varphi_{\text{1-loop
}}\mathcal Z^B_{\text{1-loop
}} \,,
\ee
where
\be\label{Zvarphi}
\mathcal Z^\varphi_{\text{1-loop
}}=\frac{1}{\mathcal Z^\varphi}\int \mathcal D\varphi_1\mathcal D\varphi_2\,e^{-\int \mathcal L^{(2)}(\varphi_1,\varphi_2)} \,, \ee
and
\be\label{Zb}
\mathcal Z^B_{\text{1-loop
}}=
\frac{1}{\mathcal Z^B}\int\mathcal D r\mathcal D\p\mathcal D B_\m \,
e^{-\int \mathcal L^{(2)}(r,\p,B_\m)}\,\,.
\ee
The partition functions  $\mathcal Z^B$ and $\mathcal Z^\varphi$ serve as normalizations and are evaluated around the trivial vacuum. Computing \eqref{Zvarphi}  yields
\be \label{spectator}
\omega^\varphi_\pm=\sqrt{J_{\ell(s)}^2+\m^2} \pm \mu\,\,.
\ee
The evaluation of the partition function \eqref{Zb} has been described in \cite{Antipin:2022hfe} and therefore, here we only highlight the main steps of the computation process. First note that the action \eqref{LquadraticU(1)} enjoys gauge invariance
\be
\label{residual}
\d r=0, \quad \d\p=v\, \a(x),\quad \d B_{\m}=-\frac{1}{g^{'}}D_{\m}\b \,,
\ee
which is expected from Elitzur's theorem \cite{Elitzur:1975im} which states that we can spontaneously break only the global part of a compact local symmetry. Hence, for a finite result we should fix the gauge. Consequently, we employ  $R_\xi$-gauge fixing 
\be\label{Zfixed}
\mathcal Z=\frac{1}{\mathcal Z^B}\int \mathcal Dr\mathcal D\p \mathcal D B_\m \,e^{-\int d^Dx \left(\mathcal L^{(2)}(r,\p,B) +\frac{1}{2}G^2\right)}\det\left(\frac{\d G}{\d\b}\right) \,,
\ee
where 
\be
\label{Gfix}
G=\frac{1}{\sqrt\xi}\left(\nabla_{\m}B^{\m}+\frac12g{'}\tilde v\,\p\right)\,,\qquad
\frac{\d G}{\d \b}=\frac{1}{g{'}\sqrt\xi}\left((-
\nabla^2)+ \frac14({g{'}}\tilde v)^2\right)\,\,.
\ee
As usual, the $\det(\d G/\d\b)$ is presented using a set of Fadeev-Popov ghosts $\bar c,c$ 
\be
\label{ghost}
\det\left(\frac{\d G}{\d \b}\right)=\int\mathcal D\bar c\,\mathcal Dc\,\,e^{-\int d^Dx\,\,\bar c\,(-\nabla^2+\frac14(g{'}v)^2)\,c}\,\,.
\ee
In addition, we split the gauge field $B_i$ in a transverse $\mathcal B_i$ and a longitudinal  $C_i$ components
\be\label{mathcalB}
B_{i}=\mathcal B_i+C_i \,,
\ee
where $\nabla_i \mathcal B^i=0$ and $C_i=\nabla_i f$ for $f$ a scalar function (for more details see Appendix \ref{sphere}). Expanding the action in eq. \eqref{Zfixed} in terms of eq. \eqref{mathcalB}, the part containing the $\mathcal B_i$ field decouples and contributes to independent dispersion relations. Performing the integration with respect to the fluctuating fields for the gauge-fixed action we obtain
\be
\int\mathcal Dr\mathcal D \p \mathcal D B_0\mathcal D C_i \,\,
e^{-\mathcal L_{fix}(r,\p,B_0,C_i)}
=\left(\det \mathcal U\right)^{1/2} \,,
\ee 
where $\mathcal U$ stands for the matrix 
\be
\mathcal U
\footnotesize
=\left(
  \begin{array}{cccc}
   - \omega^2+J_{\ell(s)}^2+2(\m^2-m^2) & -2i\m\omega   &-ig{'}v\m &0 \\
  2i\m \omega & -\omega^2+J_{\ell(s)}^2+\frac{1}{4\xi}(g{'}\tilde v)^2 & -\frac12g{'}v\left(1-\frac{1}{\xi}\right)\omega & -\frac{i}{2} g{'}v\left(1-\frac{1}{\xi}\right)|J_{\ell(s)}|\\
   -ig{'}v\m   & \frac12g{'}v\left(1-\frac{1}{\xi}\right)\omega & -\frac{1}{\xi}\omega^2+J_{\ell(s)}^2+\frac14(g{'}\tilde v)^2/4 & i\left(1-\frac{1}{\xi}\right) \omega|J_{\ell(s)}|\\
   0     & \frac{i}{2}g{'}v\left(1-\frac{1}{\xi}\right) |J_{\ell(s)}|&
 i\left(1-\frac{1}{\xi}\right)\omega|J_{\ell(s)}| &-\omega^2+\frac{1}{\xi}J_{\ell(s)}^2+\frac14(g{'}\tilde v)^2 \\ 
    \end{array}
\right)\,\,.
\ee
\normalsize
We can now obtain  the determinant in the following factorized form
\be
\label{xidet}
\xi\,\det\mathcal U=(\omega^2+\omega_+^2)(\omega^2+\omega_-^2)(\omega^2+\omega_1^2)^2 \,\,.
\ee
The detailed expressions of the dispersion relations for all fields are given in Table \ref{table1}.
\begin{table}
\label{table1}
\begin{center}\begin{tabular}{ccccc}\hline\hline
Field & $d_\ell$ & $\omega_\ell$ &  $\ell_0$ \\
\hline
$\mathcal B_i$ & $n_A(\ell)$ &  $\sqrt{J_{\ell(v)}^2 +(D-2) + \frac{1}{4}(g^{'} \tilde{v})^2}$ &1 \\   
$C_i$ & $n_b(\ell)$ &  $\sqrt{J_{\ell(s)}^2+ \frac{1}{4}(g^{'} \tilde{v})^2}$ &1 \\   
$(c,\bar c)$ & $-2 n_b(\ell)$ & $\sqrt{J_{\ell(s)}^2+ \frac{1}{4}(g^{'} \tilde{v})^2}$ &0 \\ 
$B_0$ &  $n_b(\ell)$ &  $\sqrt{J_{\ell(s)}^2+ \frac{1}{4}(g^{'} \tilde{v})^2}$ & 0\\ 
$\phi$ & $n_b (\ell)$  
& $\sqrt{J_{\ell(s)}^2+3 \mu ^2-m^2 +\frac{1}{8}(g^{'} \tilde{v})^2 \pm \sqrt{\left(3 \mu ^2-m^2-\frac{1}{8}(g^{'} \tilde{v})^2\right)^2+4 J_{\ell(s)}^2 \mu ^2}}$ & 0\\
\hline\hline
\end{tabular}\end{center}
\caption{\small The fields, their degeneracy and dispersion relations in the bosonic sector of the theory at fixed hypercharge
  }
\end{table}
Note that $\xi$ will be canceled with the evaluation of $\mathcal Z_B$ in eq. \eqref{Zb} in the denominator and we arrive at a gauge invariant result. Of course, this is expected for the computation of a physical object as is the scaling dimension of an operator. 

The bosonic contribution to $\Delta_0$ is then obtained by plugging the dispersion relations in Table \ref{table1} plus the ones given in eq.\eqref{spectator} into the general expression eq.\eqref{general0} and computing the sum over $\ell$. By proceeding as in the previous sections, we arrive at
\begin{align} \label{NLOBOS}
  \Delta_0^{(bos)} &= \frac{1}{128 \lambda ^2}\left[9 g^{\prime 4} \left(5+2 \mu ^2-7 \mu ^4\right)+12 g^{\prime 2}\lambda  \left(5-14 \mu ^2+9 \mu ^4\right)-16 \lambda^2\left(3  \mu ^2+1 \right)^2\right] \nonumber \\ &+\frac{1}{2} \sqrt{6 \mu ^2-2}+\mu + \frac{1}{2} \sum_{\ell = 1} \sigma^{(bos)}(\ell)  \,,
\end{align}
with 
\begin{align} \label{SUMBOS}
    \sigma^{(bos)}(\ell) &= \frac{1}{32 \lambda ^2 \ell}\Big[27 g^{\prime 4} \left(\mu ^2-1\right)^2-36 g^{\prime 2} \lambda  \left(\mu ^2-1\right) \left(\mu ^2+2 \ell (\ell+1)-1\right) \nonumber \\ &+16 \lambda ^2 \left(3 \mu ^4-6 \mu ^2-6 \ell (\ell+1) \mu ^2-2 \ell (\ell+1) (6 \ell (\ell+2)+1)+3\right)\Big]+(\ell+1)^2   \nonumber \\ & \times \Bigg[\sqrt{\frac{3 g^{\prime 2}\left(\mu ^2-1\right)}{4 \lambda }-\sqrt{\left(\frac{3 g^{\prime 2} \left(\mu ^2-1\right)}{4 \lambda }-3 \mu ^2+1\right)^2+4 \ell (\ell+2) \mu ^2}+3 \mu ^2+\ell (\ell+2)-1} \nonumber \\ & +\sqrt{\frac{3 g^{\prime 2}\left(\mu ^2-1\right)}{4 \lambda }+\sqrt{\left(\frac{3 g^{\prime 2} \left(\mu ^2-1\right)}{4 \lambda }-3 \mu ^2+1\right)^2+4 \ell (\ell+2) \mu ^2}+3 \mu ^2+\ell (\ell+2)-1}\nonumber \\ &+2 \sqrt{\mu ^2+\ell (\ell+2)}\Bigg] +2 \ell (\ell+2) \sqrt{\frac{3 g^{\prime 2} \left(\mu ^2-1\right)}{2 \lambda }+\ell (\ell+2)+1}  \,.
\end{align}

\section{Explicit perturbative results and discussion}
\label{disco}

We are now ready to combine our findings to check them against known perturbative results and provide novel predictions to all orders in the couplings. We start by highlighting the relevant expressions  which, in the double scaling limit defined in  eq.\eqref{dsl},  at the leading order is given in eq.\eqref{LEAD}  and at the next-to-leading  in \eqref{NLOFERM}, \eqref{SUMFERM}, \eqref{NLOBOS}, and \eqref{SUMBOS}. These  overall contribute to the scaling dimension $\Delta_Q$ of the lowest-lying scalar operators with fixed hypercharge $Q$.  We now move to compare our infinite orders computations with known perturbative results.  

\subsection{Three-loop results for $g=0$}

In the $g \to 0$ limit we can compare our findings to the anomalous dimensions of the operators defined in eq.\eqref{order} which, in the Landau gauge,  become  eq.\eqref{tortellini}. At the three-loop order, we obtain
\begin{align} \label{comparo}
    \Delta_Q &= Q + \Bigg\{\red{\frac{1}{3}\lambda Q^2}+\Bigg[\tad{ N \mathcal{Y}_u+ N \mathcal{Y}_d +  \mathcal{Y}_l}{\color{orange}{-\frac{3}{4} g^{\prime 2} - \frac{\lambda}{3}}}\Bigg] Q   \Bigg\}- \Bigg\{\red{\frac{2}{9} \lambda ^2 Q^3} -\Bigg[\tad{2 N \mathcal{Y}_{uu}+2 N \mathcal{Y}_{dd}+2 \mathcal{Y}_{ll} }\nonumber \\ &\tad{ -\frac{2}{3} \lambda  ( N \mathcal{Y}_u+ N \mathcal{Y}_d +   \mathcal{Y}_l)} {\color{orange}{-\frac{1}{3} \lambda g^{\prime 2}+\frac{g^{\prime 4}}{16}+\frac{\lambda ^2}{9}}}\Bigg] Q^2  + C_{22} Q\Bigg\}+ \Bigg\{ \red{\frac{8 }{27} \lambda ^3 Q^4} + \Bigg[ {\color{orange}{\frac{1}{16} g^{\prime 6} (9 \zeta (3)-1) }}\nonumber \\ & {\color{orange}{-\frac{1}{6} g^{\prime 4}\lambda (1 +3 \zeta (3))    +\frac{1}{3} g^{\prime 2} \lambda ^2 (3-2 \zeta (3))+\frac{4}{27} \lambda ^3 (9 \zeta (3)-8)}} \tad{+\frac{4}{27} \left( 3 N\left(\lambda ^2 \mathcal{Y}_{u}-3 \lambda  \mathcal{Y}_{uu} \right. \right. }\nonumber \\ & \tad{\left. \left. +9 \zeta (3) (\lambda  \mathcal{Y}_{uu}-2 \mathcal{Y}_{uuu})\right)   + 3 N\left(\lambda ^2 \mathcal{Y}_{d}-3 \lambda  \mathcal{Y}_{dd}+9 \zeta (3) (\lambda  \mathcal{Y}_{dd}-2 \mathcal{Y}_{ddd})\right)  + 3 \left(\lambda ^2 \mathcal{Y}_{l}-3 \lambda  \mathcal{Y}_{ll} \right. \right.} \nonumber \\ & \tad{\left. \left.+9 \zeta (3) (\lambda  \mathcal{Y}_{ll}-2 \mathcal{Y}_{lll})\right) \right)}\Bigg]  Q^3  +  C_{23} Q^2 + C_{33} Q \Bigg\}  +\cO\left(\kappa_I^4 Q^5 \right)  \,.
\end{align}
The curly brackets separate different loop orders while red, blue, and orange colors highlight the terms stemming from the small $\kappa_I Q$ expansion of $\Delta_{-1}$, $\Delta_{0}^{(fm)}$, and $ \Delta_0^{(bos)}$, respectively \footnote{While at two loops and higher the small $\kappa_I Q$ expansion of $\Delta_0$ yields directly the perturbative results, the one-loop contribution deserves more attention. In fact, the leading order of the small $\kappa_I Q$ expansion of $\Delta_0$ is a rational function of the couplings that need to be converted into the polynomial one-loop term. This is achieved by first determining the result at the fixed point as a function of $\epsilon$ and then inferring the polynomial in the coupling constants yielding the same result.}. The above has been checked via an independent diagrammatic calculation \cite{sasha}. We used the coefficients $C_{kl}$ introduced in eq.\eqref{PT} to write the terms that carry subleading powers of $Q$ and, therefore, have not been computed in the present work, since they appear at higher orders of the semiclassical expansion \eqref{ERexpansion}. However, as explained in Sec.~\ref{review}, these terms may be fixed by requiring consistency between the general structure \eqref{PT} and known perturbative results for fixed values of $Q$. In our case, the only anomalous dimension known in the literature for the family of operators eq.\eqref{tortellini} is the  one of the Higgs field corresponding to $Q=1$ \cite{Bednyakov:2013eba, Chetyrkin:2012rz, Bednyakov:2013cpa}. By combining our results with this information  we can fix the following coefficient 
\be
C_{22} =  20 g_s^2 (\mathcal{Y}_{d}+\mathcal{Y}_{u})+\frac{5 \lambda ^2}{18}+\frac{2}{3} \lambda  (3 \mathcal{Y}_{d}+3 \mathcal{Y}_{u}+\mathcal{Y}_{l})-\frac{1}{4} (51 \mathcal{Y}_{uu}+51 \mathcal{Y}_{dd}-6 \mathcal{Y}_{ud} +17 \mathcal{Y}_{ll})  \,,
\ee
where
\be
\mathcal{Y}_{ud} =  (4\pi)^4\Tr Y_u Y_u^\dagger Y_d Y_d^\dagger \ , 
\ee
allowing us to arrive, for the first time, at the full two-loop anomalous dimension of the Higgs family of operators for arbitrary $Q$. To ease the comparison with future diagrammatic computations we provide additional explicit results up to six loops in App.\ref{C}. 

We conclude this section by observing that all the terms scaling as $g^\alpha {g^{\prime}}^{\beta} \lambda^{\delta} Q^{\alpha + \beta +\delta}$ with $\beta>0$ in the anomalous dimension of the family of operators \eqref{tortellini} can be obtained from $\Delta_0^{(bos)}$ in eq.\eqref{NLOBOS} by replacing ${g^{\prime}}^{2} \rightarrow {g^{\prime}}^{2} + g^2$. This can be traced back to the absence of mixing between electrically charged and neutral modes when calculating the quadratic Lagrangian in the bosonic sector.

\subsection{$\Delta_Q$ for $g \neq 0$}

We now study the scaling dimension $\Delta_{Q}$ for non-vanishing $SU(2)_L$ gauge coupling $g$. We start by solving eq.\eqref{theequation} in the limit of small 't Hooft-like couplings. To this end, we note that only two out of the six solutions of eq.\eqref{theequation}, feature the correct $g \to 0$ limit. These are
\begin{align}
    \mu_0^{(\pm)} & = 1 \pm \frac{3 g }{2 \sqrt{2}}\sqrt{Q} + Q \left(\frac{2 \lambda }{3}-\frac{g^2}{8}\right) \pm  \frac{5 g  \left(3 g^2-64 \lambda \right)}{384 \sqrt{2}} Q^{3/2}+ \left(-\frac{3 g^4}{512}+\frac{3 g^2 \lambda }{8}-\frac{2 \lambda ^2}{3}\right) Q^2 \nonumber \\ & \pm\frac{7 g}{16384 \sqrt{2}}\left(3 g^4-640 g^2 \lambda +4096 \lambda ^2\right)  Q^{5/2} + \frac{1}{108} \lambda  \left(9 g^4-144 g^2 \lambda +128 \lambda ^2\right)  Q^3 + \mathcal{O}\left( Q^{7/2} \right) \,, 
\end{align}
and lead to the following expressions for $\Delta_{-1}$
\begin{align}
\Delta_{-1}^{(\pm)} &=Q \pm \frac{g Q^{3/2}}{\sqrt{2}}  + \left(\frac{\lambda }{3}-\frac{g^2}{16}\right) Q^2 \pm \frac{g}{192 \sqrt{2}} \left(3 g^2 - 64 \lambda \right) Q^{5/2} \nonumber \\ &  + \left(-\frac{g^4}{512}+\frac{g^2 \lambda }{8}-\frac{2 \lambda ^2}{9}\right) Q^3+ \mathcal{O}\left( Q^{7/2} \right) \,.
\end{align}
The free energy in  eq.\eqref{LEAD} is a unique function of the chemical potential $\mu$ while 
the apparent multivaluedness of $\Delta_{-1}$ arises via a naive definition of the Legendre transform. To resolve this issue we recall that for a real-valued function $f(x)$ of a real variable $x$, the standard definition of the Legendre transform is
\be
L[f](y) = \underset{x}{\text{sup}}(x y - f(x))  \,,
\ee
that naturally selects the lowest energy state. While at the classical level, the theory is conformal for arbitrary values of $g$ and $\lambda$, the Wilson-Fisher fixed point occurs at complex values of the couplings defining a pair of complex CFTs with complex conjugate CFT data \cite{Gorbenko:2018ncu}. In this case, the problem of defining the Legendre transform is not immediate due to the presence of complex saddles. This issue is addressed via the Picard-Lefschetz theory \cite{Witten:2010cx, Cherman:2014ofa, Tanizaki:2014xba} and we plan to revisit it in the future. However, we observe that the  average over the two solutions, i.e.
\be
\Delta_{-1}^{(\text{Avg})} =\frac12 \left(\Delta_{-1}^{(+)} + \Delta_{-1}^{(-)} \right) =Q  + \left(\frac{\lambda }{3}-\frac{g^2}{16}\right) Q^2+ \left(-\frac{g^4}{512}+\frac{g^2 \lambda }{8}-\frac{2 \lambda ^2}{9}\right)Q^3 + \mathcal{O}\left( Q^{4} \right)  \,,
\ee
uniquely yields the correct $g \to 0$ limit without featuring unexpected  non-integer powers of $Q$.

\vskip 1cm
We provided the first determination of the anomalous dimensions of the family of composite operators made out of Higgs fields with fixed hypercharge at all orders in the SM couplings and to leading and subleading orders in the charge. Future directions include a more detailed identification of the composite operators, the computation of $\Delta_0$ at nonvanishing weak gauge coupling, and the integration of our results into the computation of the multi-Higgs production \cite{Goldberg:1990qk, Rubakov:1995hq, Degrande:2016oan} for colliders applications.

\section*{Acknowledgments}
The work of J.B. was supported by the World Premier International Research Center Initiative (WPI Initiative), MEXT, Japan; and also supported by the JSPS KAKENHI Grant Number JP23K19047. The work of P. P is supported  by the National Research Foundation of Korea (NRF) grant funded by the Korea government (MSIT) (No. 2023R1A2C1006975) and from the JRG Program at the APCTP through the Science and Technology Promotion Fund and Lottery Fund of the Korean Government.  The work of F.S.~is partially supported by the Carlsberg Foundation, grant CF22-0922.  We thank Alexander Bednyakov for sharing with us his three-loops perturbative results for the Standard Model anomalous dimensions of the family of Higgs operators with hypercharge $Q$.

\appendix

\section{One-loop SM $\beta$ functions} 
\label{appendicite}

Below we collect the one-loop SM beta functions needed for this work:
\begin{align}
    \beta_{Y_u} &= -\frac{9g ^2}{4}-\frac{17 g^{\prime 2}}{20}-8 g_S^2+3 (\mathcal{Y}_d+\mathcal{Y}_u)+\mathcal{Y}_l+\frac{3 (Y_u-Y_d)}{2}   \,,\\
       \beta_{Y_d} &= -\frac{9 g^2}{4}-\frac{g^{\prime 2}}{4}-8 g_S^2+3 (\mathcal{Y}_d+\mathcal{Y}_u)+\mathcal{Y}_l+\frac{3 (Y_d-Y_u)}{2}  \,,\\
          \beta_{Y_l} &= -\frac{9g ^2}{4} -\frac{9 g^{\prime 2}}{4}+3 (\mathcal{Y}_d+\mathcal{Y}_u)+\mathcal{Y}_l+\frac{3 Y_l}{2}  \,,\\
            \beta_\l &=  -12 \left(N \mathcal{Y}_{uu} + N \mathcal{Y}_{dd}  + \mathcal{Y}_{ll} \right)+4  \lambda \left( N  \mathcal{Y}_{u}+ N   \mathcal{Y}_{d} +  \mathcal{Y}_{l}\right) \nonumber \\ & +\frac{9 g^{\prime 4}}{4}+\frac{9 g^{\prime 2} g^2}{2}-3 g^{\prime 2} \lambda +\frac{27 g^4}{4}-9 g^2 \lambda +4 \lambda ^2   \,, \\ 
  \beta_{g^\prime} &= \frac{41 g^{\prime 3}}{6}  \,, \qquad
   \beta_{g} = -\frac{19}{6} g^3  \,,
\end{align}
where the beta functions for the Yukawa matrix couplings are defined as
\be
M \frac{d}{d M} Y_f = \beta_{Y_f} Y_f  \,, \qquad M \frac{d}{d M} Y_f^\dagger = Y_f^\dagger  \beta_{Y_f^\dagger } \,, \qquad f= u, d, l \,,
\ee
with $M$ the RG scale. At the one-loop level we have $ \beta_{Y_f}  = \beta_{Y_f^\dagger }$.

\section{Scalar, Gauge and Spinor fields on $S^{D-1}$}
\label{sphere}
 Here, we provide details about the Laplacian for scalar, vector, and spinor fields on $S^{D-1}$. Although in the main text we set the radius of the $S^{D-1}$ sphere to one here we re-introduce and define it as $R$. Therefore, the conformally invariant action for a real scalar field $H$ on a curved manifold $\mathcal M$ with Ricci scalar $\mathcal R$ is

\be
S=\frac{1}{2}\int d^{D-1}x\sqrt{-g}\,H \left(-\nabla^2_{S^{D-1} }+\frac{D-2}{4(D-1)}\mathcal R\right) H \,,
\ee
where $\nabla^2\equiv\nabla_{\m}\nabla^{\m}$.  The Laplacian $\D$ of a scalar field $H$ on curved background $\mathcal M$ is defined as
\be
\Delta H=-\frac{1}{\sqrt{g}}\del_{\m}\left(\sqrt{-g}g^{\m\n}\del_{\n} H \right)=-\nabla^2_{\mathcal M} H \,.
\ee

The eigenvectors of the scalar Laplacian on $S^{D-1}$ are given by the spherical harmonics $Y_{\ell}$. These are labeled by the angular momentum quantum number, $\ell\in \mathbb{Z}\geq0$. The eigenvalue equation reads
\be
\label{eigenscalar}
-\nabla^2_{S^D}\,Y_{\ell}=J^2_{\ell (s)} Y_{\ell} \,,
\ee
where the eigenvalues are given by
\be
J^2_{\ell (s)}  = \frac{1}{R^2}\ell(\ell +D-2) \,,
\ee
and have degeneracy 
\be
n_b(\ell)=\frac{(2\ell+D-2)\G(\ell+D-2)}{\G(D-1)\G(\ell+1)}\,\,.
\ee
Proceeding with  gauge bosons, the pure Maxwell theory on the sphere reads
\be
\begin{split}
S(A)=&\frac{1}{4}\int d^{D-1}x\sqrt{-g}\,F_{\m\n}F^{\m\n}\\
=&
\frac{1}{2}\int d^{D-1}x\sqrt{-g}\,A^{\n}\left(-\d^{\m}_{\n}\,\nabla^2+\mathcal R^{\m}_{\n}+\nabla_{\n}\nabla^{\m}\right)A_{\m} \,,
\end{split}
\ee
where in the last step we used the fact that $[\nabla^{\m},\nabla_{\n}]A_{\mu}=\mathcal R^{\m}_{\n}A_{\m}$ with  $\mathcal R^{\m}_{\n}=\frac{D-2}{R^2}\,\d^{\m}_{\n}$ is the Ricci tensor on $S^{D-1}$.
For the component $A_0$ which is a scalar on $S^{D-1}$ the Ricci tensor does not contribute and $\nabla^2$ is equivalent to the scalar Laplacian. 
The vector field, as usual, can be decomposed as the kernel plus the image of $\nabla^i$, namely  $A^i=B^i+C^i$ with
\be
\label{decomp}
\nabla_iB^i=0,\qquad C^i=\nabla^if \,,
\ee
where $f$ is an arbitrary function. These two modes have different eigenvalues when Laplacian acts on them. In particular, $f$ is a scalar while 
the $B^i$ modes are spanned by $V_{\ell}^i$ labelled by $\ell\in Z>0$ and satisfying
\be
\label{eigenvec}
-\nabla^2_{S^D} V_{\ell}^i=J_{\ell(v)}^2 V_{\ell}^i \,,
\ee
where the eigenvalues read
\be
J_{\ell(v)}^2=\frac{1}{R^2}\left(\ell(\ell+D-2)-1\right)\,\,,
\ee
and have the following degeneracy
\be
n_{A}(\ell)=\frac{\ell(\ell+D-2)(2\ell+D-2)\G(\ell+D-3)}{\G(\ell+2)\G(D-2)}\,\,.
\label{nadegeneracy}
\ee
We conclude with the action of the Laplacian on $S^{D-1}$ on spinor fields yields the following eigenvalues and degeneracy 
\be \label{eigenforferm}
\l_{f\pm}(\ell)=\pm \frac{1}{R}\left(\ell+\frac{D-1}{2}\right),\qquad 
 n_f(\ell)=\frac{4\,\G(\ell+D-1)}{\G(D-1)\G(\ell+1)}\,\,.
\ee

\section{Explicit $6$-loops results for the anomalous dimensions}
\label{C}
Here we provide explicit results for $\Delta_Q$ in the $g \to 0$ limit. We use the notation introduced in eq.\eqref{PT} and list the values of the coefficients $C_{0l}$ and $C_{1l}$ for $l=4,5,6$ while the coefficients with $l < 4$ can be read off from eq.\eqref{comparo}. 

The $C_{0l}$ coefficients appear in the small 't Hooft-like coupling expansion of $\Delta_{-1}$ and are given by
\begin{align} 
		C_{04}=-\frac{14}{27}  \lambda ^4\,,\quad   C_{05}= \frac{256}{243}\lambda ^5\,,\quad   C_{06}= -\frac{572 }{243}\lambda ^6 \,.
\end{align}

Analogously, the $C_{1l}$ coefficients stem from the small 't Hooft-like coupling expansion of $\Delta_{0}$ and read

\begin{align} 	
  C_{14}&= \frac{1}{81} (6N\mathcal{F}_4^{(u)} + 6N\mathcal{F}_4^{(d)} + 6\mathcal{F}_4^{(l)} -2 \lambda ^4 (89 \zeta (3)+85 \zeta (5)-138))\nonumber \\ &+\frac{1}{128} g^{\prime 8} (41 \zeta (3)-125 \zeta (5)+17)-\frac{1}{48} g^{\prime 6} \lambda  (134 \zeta (3)-140 \zeta (5)+25)\nonumber \\ &+\frac{1}{24} g^{\prime 4} \lambda ^2 (80 \zeta (3)-80 \zeta (5)+33)-\frac{2}{27} g^{\prime 2} \lambda ^3 (7 \zeta (3)-40 \zeta (5)+37) \,, 
  \end{align}
  \begin{align}
   C_{15}&= \frac{1}{62208}  \Bigg[512\left(3 N \mathcal{F}_5^{(u)} +3 N \mathcal{F}_5^{(d)} +3 \mathcal{F}_5^{(l)}  +2 \lambda ^5 (280 \zeta (3)+260 \zeta (5)+231 \zeta (7)-579)\right)  \nonumber \\ &   -243 g^{\prime 10} (224 \zeta (3)-50 \zeta (5)-441 \zeta (7)+95)  +648 g^{\prime 8} \lambda  (413 \zeta (3)-50 \zeta (5)-525 \zeta (7)+167) \nonumber \\ & -864 g^{\prime 6} \lambda ^2 (164 \zeta (3)-315 \zeta (5)-140 \zeta (7)+152)-576 g^{\prime 4} \lambda ^3 (368 \zeta (3)+7 (60 \zeta (5)-80 \zeta (7)+31)) \nonumber \\ &   +768 g^{\prime 2} \lambda ^4 (156 \zeta (3)-130 \zeta (5)-560 \zeta (7)+559)\Bigg]  \,, 
  \end{align}
  \begin{align}
   C_{16}&= \frac{1}{729} \left(6 N \mathcal{F}_6^{(u)} + 6 N \mathcal{F}_6^{(d)} + 6 \mathcal{F}_6^{(l)}    -2 \lambda ^6 (3934 \zeta (3)+3522 \zeta (5)+3500 \zeta (7) +2730 \zeta (9)-9935)\right)   \nonumber \\ &    +\frac{g^{\prime 12} }{2048}(5662 \zeta (3)+462 \zeta (5)-3514 \zeta (7)       -7434 \zeta (9)+2487) \nonumber \\ &+\frac{1}{192} g^{\prime 10} \lambda  (-2956 \zeta (3)+319 \zeta (5)+1680 \zeta (7)+2583 \zeta (9)-1270) \nonumber \\ & +\frac{1}{576} g^{\prime 8} \lambda ^2 (14932 \zeta (3)-8350 \zeta (5)-10185 \zeta (7)-6300 \zeta (9)+6286) \nonumber \\ & +\frac{1}{108} g^{\prime 6} \lambda ^3 (-2413 \zeta (3)+1335 \zeta (5)+910 \zeta (7)+840 \zeta (9)-650) \nonumber \\ &  +\frac{1}{162} g^{\prime 4} \lambda ^4 (2645 \zeta (3)+515 \zeta (5)-490 \zeta (7)-2520 \zeta (9)+1421)\nonumber \\ & +\frac{1}{243} g^{\prime 2} \lambda ^5 (-1886 \zeta (3)+46 \zeta (5)+2464 \zeta (7)+4032 \zeta (9)-4633)  \,,
\end{align}
where
\begin{align}
 \mathcal{F}_4^{(f)} &=   45 \zeta (5) \left(\mathcal{Y}_{f}^4-6 \mathcal{Y}_{f}^2 \mathcal{Y}_{ff}+8 \mathcal{Y}_{f} \mathcal{Y}_{fff}+3 \mathcal{Y}_{ff}^2-4 \lambda  \mathcal{Y}_{fff}  \right) \nonumber \\  & -2 \lambda  \zeta (3) \left(\lambda ^2 \mathcal{Y}_{f}+45 \lambda  \mathcal{Y}_{ff}-108 \mathcal{Y}_{fff} \right) +24 \lambda ^2 \mathcal{Y}_{ff} -8 \lambda ^3 \mathcal{Y}_{f}  \,,
\end{align}
\begin{align}
   \mathcal{F}_5^{(f)} &= 63 \zeta (7) \left(5 \lambda  \left(\mathcal{Y}_{f}^4-6 \mathcal{Y}_{f}^2 \mathcal{Y}_{ff}+8 \mathcal{Y}_{f} \mathcal{Y}_{fff}+3 \mathcal{Y}_{ff}^2\right)-6 \left(\mathcal{Y}_{f}^5-5 \mathcal{Y}_{f}^3 \mathcal{Y}_{ff}   +5 \mathcal{Y}_{f}^2 \mathcal{Y}_{fff}+5 \mathcal{Y}_{ff} \mathcal{Y}_{fff}\right)\right)   \nonumber \\ & + 60 \lambda  \zeta (5) \left(-6 \left(\mathcal{Y}_{f}^4-6 \mathcal{Y}_{f}^2 \mathcal{Y}_{ff}+8 \mathcal{Y}_{f} \mathcal{Y}_{fff}+3 \mathcal{Y}_{ff}^2\right)+\lambda ^2 \mathcal{Y}_{ff}+18 \lambda  \mathcal{Y}_{fff}\right)   \nonumber \\ &+4 \lambda ^2 \zeta (3) \left(5 \lambda ^2 \mathcal{Y}_{f}+144 \lambda  \mathcal{Y}_{ff}-378 \mathcal{Y}_{fff}\right)-126 \lambda ^3 \mathcal{Y}_{ff} +42 \lambda ^4 \mathcal{Y}_{f} \,,
\end{align}
\begin{align}
    \mathcal{F}_6^{(f)} &=1701 \zeta (9) \left(\mathcal{Y}_{f}^6-3 \mathcal{Y}_{f}^4 \mathcal{Y}_{ff}+4 \mathcal{Y}_{f}^3 \mathcal{Y}_{fff}-9 \mathcal{Y}_{f}^2 \mathcal{Y}_{ff}^2   -2 \lambda  \left(\mathcal{Y}_{f}^5-5 \mathcal{Y}_{f}^3 \mathcal{Y}_{ff}+5 \mathcal{Y}_{f}^2 \mathcal{Y}_{fff}+5 \mathcal{Y}_{ff} \mathcal{Y}_{fff}\right)  \right.  \nonumber \\ &   \left. +12 \mathcal{Y}_{f} \mathcal{Y}_{ff} \mathcal{Y}_{fff}+3 \mathcal{Y}_{ff}^3+4 \mathcal{Y}_{fff}^2\right)-315 \lambda  \zeta (7) \left(7 \lambda  \left(\mathcal{Y}_{f}^4-6 \mathcal{Y}_{f}^2 \mathcal{Y}_{ff}+8 \mathcal{Y}_{f} \mathcal{Y}_{fff}+3 \mathcal{Y}_{ff}^2\right)  \right.  \nonumber \\ &   \left.  -12 \left(\mathcal{Y}_{f}^5-5 \mathcal{Y}_{f}^3 \mathcal{Y}_{ff}+5 \mathcal{Y}_{f}^2 \mathcal{Y}_{fff}+5 \mathcal{Y}_{ff} \mathcal{Y}_{fff}\right)  +4 \lambda ^2 \mathcal{Y}_{fff}\right) -2 \lambda ^2 \zeta (5) \left(-1440 \left(\mathcal{Y}_{f}^4-6 \mathcal{Y}_{f}^2 \mathcal{Y}_{ff} \right.  \right.  \nonumber \\ &   \left. \left.  +8 \mathcal{Y}_{f} \mathcal{Y}_{fff}+3 \mathcal{Y}_{ff}^2\right)  +\lambda ^3 \mathcal{Y}_{f}+345 \lambda ^2 \mathcal{Y}_{ff}   +3600 \lambda  \mathcal{Y}_{fff}\right)   -2 \lambda ^3 \zeta (3) \left(89 \lambda ^2 \mathcal{Y}_{f}+2079 \lambda  \mathcal{Y}_{ff}-5760 \mathcal{Y}_{fff}\right)  \nonumber \\ &   +768 \lambda ^4 \mathcal{Y}_{ff} -256 \lambda ^5 \mathcal{Y}_{f}   \,.
\end{align}


\normalsize
 
\begin{thebibliography}{1}
\small

\bibitem{Hellerman:2015nra}
S.~Hellerman, D.~Orlando, S.~Reffert and M.~Watanabe,
``On the CFT Operator Spectrum at Large Global Charge,''
JHEP \textbf{12} (2015), 071
doi:10.1007/JHEP12(2015)071
[arXiv:1505.01537 [hep-th]].


\bibitem{Banerjee:2017fcx}
D.~Banerjee, S.~Chandrasekharan and D.~Orlando,
``Conformal dimensions via large charge expansion,''
Phys. Rev. Lett. \textbf{120} (2018) no.6, 061603
doi:10.1103/PhysRevLett.120.061603
[arXiv:1707.00711 [hep-lat]].



\bibitem{Orlando:2019hte}
D.~Orlando, S.~Reffert and F.~Sannino,
``A safe CFT at large charge,''
JHEP \textbf{08} (2019), 164
doi:10.1007/JHEP08(2019)164
[arXiv:1905.00026 [hep-th]].

\bibitem{Orlando:2019skh}
D.~Orlando, S.~Reffert and F.~Sannino,
``Near-Conformal Dynamics at Large Charge,''
Phys. Rev. D \textbf{101} (2020) no.6, 065018
doi:10.1103/PhysRevD.101.065018
[arXiv:1909.08642 [hep-th]].


\bibitem{Gaume:2020bmp}
L.~\'A.~Gaum\'e, D.~Orlando and S.~Reffert,
``Selected topics in the large quantum number expansion,''
Phys. Rept. \textbf{933} (2021), 1-66
doi:10.1016/j.physrep.2021.08.001
[arXiv:2008.03308 [hep-th]].


\bibitem{Orlando:2020yii}
D.~Orlando, S.~Reffert and F.~Sannino,
``Charging the Conformal Window,''
Phys. Rev. D \textbf{103} (2021) no.10, 105026
doi:10.1103/PhysRevD.103.105026
[arXiv:2003.08396 [hep-th]].

\bibitem{Hellerman:2021qzz}
S.~Hellerman, D.~Orlando, V.~Pellizzani, S.~Reffert and I.~Swanson,
``Nonrelativistic CFTs at large charge: Casimir energy and logarithmic enhancements,''
JHEP \textbf{05} (2022), 135
doi:10.1007/JHEP05(2022)135
[arXiv:2111.12094 [hep-th]].

\bibitem{Dondi:2022zna}
N.~Dondi, S.~Hellerman, I.~Kalogerakis, R.~Moser, D.~Orlando and S.~Reffert,
``Fermionic CFTs at large charge and large N,''
JHEP \textbf{08} (2023), 180
doi:10.1007/JHEP08(2023)180
[arXiv:2211.15318 [hep-th]].

\bibitem{Hellerman:2023myh}
S.~Hellerman, D.~Krichevskiy, D.~Orlando, V.~Pellizzani, S.~Reffert and I.~Swanson,
``The unitary Fermi gas at large charge and large N,''
[arXiv:2311.14793 [hep-th]].

\bibitem{Antipin:2020abu}
O.~Antipin, J.~Bersini, F.~Sannino, Z.~W.~Wang and C.~Zhang,
``Charging the $O(N)$ model,''
Phys. Rev. D \textbf{102} (2020) no.4, 045011
doi:10.1103/PhysRevD.102.045011
[arXiv:2003.13121 [hep-th]].

\bibitem{Antipin:2020rdw}
O.~Antipin, J.~Bersini, F.~Sannino, Z.~W.~Wang and C.~Zhang,
``Charging non-Abelian Higgs theories,''
Phys. Rev. D \textbf{102} (2020) no.12, 125033
doi:10.1103/PhysRevD.102.125033
[arXiv:2006.10078 [hep-th]].

\bibitem{Antipin:2022naw}
O.~Antipin, J.~Bersini and P.~Panopoulos,
``Yukawa interactions at large charge,''
JHEP \textbf{10} (2022), 183
doi:10.1007/JHEP10(2022)183
[arXiv:2208.05839 [hep-th]].

\bibitem{Antipin:2022hfe}
O.~Antipin, A.~Bednyakov, J.~Bersini, P.~Panopoulos and A.~Pikelner,
``Gauge Invariance at Large Charge,''
Phys. Rev. Lett. \textbf{130} (2023) no.2, 2
doi:10.1103/PhysRevLett.130.021602
[arXiv:2210.10685 [hep-th]].

\bibitem{Antipin:2022dsm}
O.~Antipin, J.~Bersini, F.~Sannino and M.~Torres,
``The analytic structure of the fixed charge expansion,''
JHEP \textbf{06} (2022), 041
doi:10.1007/JHEP06(2022)041
[arXiv:2202.13165 [hep-th]].

\bibitem{Monin:2016jmo}
A.~Monin, D.~Pirtskhalava, R.~Rattazzi and F.~K.~Seibold,
``Semiclassics, Goldstone Bosons and CFT data,''
JHEP \textbf{06} (2017), 011
doi:10.1007/JHEP06(2017)011
[arXiv:1611.02912 [hep-th]].

\bibitem{Badel:2019oxl}
G.~Badel, G.~Cuomo, A.~Monin and R.~Rattazzi,
``The Epsilon Expansion Meets Semiclassics,''
JHEP \textbf{11} (2019), 110
doi:10.1007/JHEP11(2019)110
[arXiv:1909.01269 [hep-th]].


\bibitem{Badel:2019khk}
G.~Badel, G.~Cuomo, A.~Monin and R.~Rattazzi,
``Feynman diagrams and the large charge expansion in $3-\varepsilon$ dimensions,''
Phys. Lett. B \textbf{802} (2020), 135202
doi:10.1016/j.physletb.2020.135202
[arXiv:1911.08505 [hep-th]].

\bibitem{Cuomo:2020rgt}
G.~Cuomo,
``A note on the large charge expansion in 4d CFT,''
Phys. Lett. B \textbf{812} (2021), 136014
doi:10.1016/j.physletb.2020.136014
[arXiv:2010.00407 [hep-th]].

\bibitem{Cuomo:2022kio}
G.~Cuomo and Z.~Komargodski,
``Giant Vortices and the Regge Limit,''
JHEP \textbf{01} (2023), 006
doi:10.1007/JHEP01(2023)006
[arXiv:2210.15694 [hep-th]].

\bibitem{Arias-Tamargo:2019xld}
G.~Arias-Tamargo, D.~Rodriguez-Gomez and J.~G.~Russo,
``The large charge limit of scalar field theories and the Wilson-Fisher fixed point at $\epsilon=0$,''
JHEP \textbf{10} (2019), 201
doi:10.1007/JHEP10(2019)201
[arXiv:1908.11347 [hep-th]].

\bibitem{Jack:2020wvs}
I.~Jack and D.~R.~T.~Jones,
``Anomalous dimensions for $\phi^n$ in scale invariant $d=3$ theory,''
Phys. Rev. D \textbf{102} (2020) no.8, 085012
doi:10.1103/PhysRevD.102.085012
[arXiv:2007.07190 [hep-th]].

\bibitem{Jack:2021lja}
I.~Jack and D.~R.~T.~Jones,
``Anomalous dimensions at large charge for U(N)\texttimes{}U(N) theory in three and four dimensions,''
Phys. Rev. D \textbf{104} (2021) no.10, 105017
doi:10.1103/PhysRevD.104.105017
[arXiv:2108.11161 [hep-th]].


\bibitem{Jafferis:2017zna}
D.~Jafferis, B.~Mukhametzhanov and A.~Zhiboedov,
``Conformal Bootstrap At Large Charge,''
JHEP \textbf{05} (2018), 043
doi:10.1007/JHEP05(2018)043
[arXiv:1710.11161 [hep-th]].


\bibitem{Caetano:2023zwe}
J.~Caetano, S.~Komatsu and Y.~Wang,
``Large Charge 't Hooft Limit of $\mathcal{N}=4$ Super-Yang-Mills,''
[arXiv:2306.00929 [hep-th]].


\bibitem{Bednyakov:2022guj}
A.~Bednyakov and A.~Pikelner,
``Six-loop anomalous dimension of the $\phi^Q$ operator in the O(N) symmetric model,''
Phys. Rev. D \textbf{106} (2022) no.7, 076015
doi:10.1103/PhysRevD.106.076015
[arXiv:2208.04612 [hep-th]].

\bibitem{Cardy:1984rp}
J.~L.~Cardy,
``Conformal invariance and universality in finite-size scaling,''
J. Phys. A \textbf{17} (1984), L385-L387

\bibitem{Aoki:2022ugd}
S.~Aoki,
``Noether\textquoteright{}s 1st theorem with local symmetries,''
PTEP \textbf{2022} (2022) no.12, 123A02
doi:10.1093/ptep/ptac160
[arXiv:2206.00283 [hep-th]].

\bibitem{Elitzur:1975im}
S.~Elitzur,
``Impossibility of Spontaneously Breaking Local Symmetries,''
Phys. Rev. D \textbf{12} (1975), 3978-3982
doi:10.1103/PhysRevD.12.3978
  
\bibitem{Dirac}
P. A. M. Dirac,
{\it The Principles of Quantum Mechanics}, 
Can. J. Phys. 33, 650 (1955);  (Oxford University Press,
Oxford, 1958), 4th Ed.


\bibitem{Caudy:2007sf}
W.~Caudy and J.~Greensite,
``On the ambiguity of spontaneously broken gauge symmetry,''
Phys. Rev. D \textbf{78} (2008), 025018
doi:10.1103/PhysRevD.78.025018
[arXiv:0712.0999 [hep-lat]].

\bibitem{Greensite:2017ajx}
J.~Greensite and K.~Matsuyama,
``Confinement criterion for gauge theories with matter fields,''
Phys. Rev. D \textbf{96} (2017) no.9, 094510
doi:10.1103/PhysRevD.96.094510
[arXiv:1708.08979 [hep-lat]].

\bibitem{sasha}
A. Bednyakov, ``Private Communication,''

\bibitem{Bednyakov:2013eba}
A.~V.~Bednyakov, A.~F.~Pikelner and V.~N.~Velizhanin,
``Higgs self-coupling beta-function in the Standard Model at three loops,''
Nucl. Phys. B \textbf{875} (2013), 552-565
doi:10.1016/j.nuclphysb.2013.07.015
[arXiv:1303.4364 [hep-ph]].

\bibitem{Bednyakov:2013cpa}
A.~V.~Bednyakov, A.~F.~Pikelner and V.~N.~Velizhanin,
``Three-loop Higgs self-coupling beta-function in the Standard Model with complex Yukawa matrices,''
Nucl. Phys. B \textbf{879} (2014), 256-267
doi:10.1016/j.nuclphysb.2013.12.012
[arXiv:1310.3806 [hep-ph]].

\bibitem{Chetyrkin:2012rz}
K.~G.~Chetyrkin and M.~F.~Zoller,
``Three-loop \textbackslash{}beta-functions for top-Yukawa and the Higgs self-interaction in the Standard Model,''
JHEP \textbf{06} (2012), 033
doi:10.1007/JHEP06(2012)033
[arXiv:1205.2892 [hep-ph]].



\bibitem{Sannino:2002wp}
F.~Sannino,
`General structure of relativistic vector condensation,''
Phys. Rev. D \textbf{67} (2003), 054006
doi:10.1103/PhysRevD.67.054006
[arXiv:hep-ph/0211367 [hep-ph]].


\bibitem{Gusynin:2003yu}
V.~P.~Gusynin, V.~A.~Miransky and I.~A.~Shovkovy,
``Spontaneous rotational symmetry breaking and roton - like excitations in gauged $\sigma$-model at finite density,''
Phys. Lett. B \textbf{581} (2004), 82-92
[erratum: Phys. Lett. B \textbf{734} (2014), 407-407]
doi:10.1016/j.physletb.2014.05.029
[arXiv:hep-ph/0311025 [hep-ph]].

\bibitem{Kapusta:1981aa}
J.~I.~Kapusta,
``Bose-Einstein Condensation, Spontaneous Symmetry Breaking, and Gauge Theories,''
Phys. Rev. D \textbf{24} (1981), 426-439
doi:10.1103/PhysRevD.24.426

\bibitem{Kapusta:1990qc}
J.~I.~Kapusta,
``Phase Diagram of Electroweak Theory,''
Phys. Rev. D \textbf{42} (1990), 919-925
doi:10.1103/PhysRevD.42.919


\bibitem{Loewe:2004zw}
M.~Loewe, S.~Mendizabal and J.~C.~Rojas,
``Weinberg-Salam model at finite temperature and density,''
Phys. Lett. B \textbf{617} (2005), 87-91
doi:10.1016/j.physletb.2005.05.010
[arXiv:hep-ph/0412392 [hep-ph]].

\bibitem{Hama:2011rt}
Y.~Hama, T.~Hatsuda and S.~Uchino,
``Higgs Mechanism with Type-II Nambu-Goldstone Bosons at Finite Chemical Potential,''
Phys. Rev. D \textbf{83} (2011), 125009
doi:10.1103/PhysRevD.83.125009
[arXiv:1102.4145 [hep-ph]].

\bibitem{Watanabe:2014qla}
H.~Watanabe and H.~Murayama,
``Spontaneously broken non-Abelian gauge symmetries in nonrelativistic systems,''
Phys. Rev. D \textbf{90} (2014) no.12, 121703
doi:10.1103/PhysRevD.90.121703
[arXiv:1405.0997 [hep-th]].

\bibitem{Gorbenko:2018ncu}
V.~Gorbenko, S.~Rychkov and B.~Zan,
``Walking, Weak first-order transitions, and Complex CFTs,''
JHEP \textbf{10} (2018), 108
doi:10.1007/JHEP10(2018)108
[arXiv:1807.11512 [hep-th]].

\bibitem{Witten:2010cx}
E.~Witten,
``Analytic Continuation Of Chern-Simons Theory,''
AMS/IP Stud. Adv. Math. \textbf{50} (2011), 347-446
[arXiv:1001.2933 [hep-th]].

\bibitem{Cherman:2014ofa}
A.~Cherman, D.~Dorigoni and M.~Unsal,
``Decoding perturbation theory using resurgence: Stokes phenomena, new saddle points and Lefschetz thimbles,''
JHEP \textbf{10} (2015), 056
doi:10.1007/JHEP10(2015)056
[arXiv:1403.1277 [hep-th]].

\bibitem{Tanizaki:2014xba}
Y.~Tanizaki and T.~Koike,
``Real-time Feynman path integral with Picard\textendash{}Lefschetz theory and its applications to quantum tunneling,''
Annals Phys. \textbf{351} (2014), 250-274
doi:10.1016/j.aop.2014.09.003
[arXiv:1406.2386 [math-ph]].

\bibitem{Goldberg:1990qk}
H.~Goldberg,
``Breakdown of perturbation theory at tree level in theories with scalars,''
Phys. Lett. B \textbf{246} (1990), 445-450
doi:10.1016/0370-2693(90)90628-J

\bibitem{Rubakov:1995hq}
V.~A.~Rubakov,
``Nonperturbative aspects of multiparticle production,''
[arXiv:hep-ph/9511236 [hep-ph]].

\bibitem{Degrande:2016oan}
C.~Degrande, V.~V.~Khoze and O.~Mattelaer,
``Multi-Higgs production in gluon fusion at 100 TeV,''
Phys. Rev. D \textbf{94} (2016), 085031
doi:10.1103/PhysRevD.94.085031
[arXiv:1605.06372 [hep-ph]].


\end{thebibliography}
\end{document}